\documentclass[sigconf,nonacm]{acmart}

\usepackage{graphicx}
\usepackage{subfiles}
\usepackage{soul,color}

\usepackage{todonotes}
\usepackage{booktabs}
\usepackage{adjustbox}
\PassOptionsToPackage{hyphens}{url}\usepackage{hyperref}

\usepackage{dsfont}
\usepackage{amsmath}
\usepackage{nicefrac}

\usepackage{amsthm}

\theoremstyle{definition}
\newtheorem{definition}{Question}

\usepackage{CJKutf8}

\usepackage{caption}
\usepackage{subcaption}
\usepackage{tablefootnote}
\usepackage{pifont}
\newcommand{\cmark}{\textcolor{teal}{\ding{51}}}%
\newcommand{\xmark}{\textcolor{red}{\ding{55}}}%
\newcommand\ml[1]{\textcolor{teal}{#1}}

\usepackage{multicol}
\usepackage{multirow}
\usepackage{url}

\usepackage{tikz}
\tikzset{
  treenode/.style = {shape=rectangle, rounded corners,
                     draw, align=center,
                     top color=white, bottom color=white},
  root/.style     = {treenode, font=\Large, bottom color=red!30},
  env/.style      = {treenode, font=\normalsize},
  dummy/.style    = {circle,draw}
}

\usepackage{xcolor}

\newcommand{\clirtask}{CLIR}

\AtBeginDocument{%
  \providecommand\BibTeX{{%
    \normalfont B\kern-0.5em{\scshape i\kern-0.25em b}\kern-0.8em\TeX}}}

\begin{document}

\title[Overview of the TREC 2024 NeuCLIR Track]{Overview of the TREC 2024 NeuCLIR Track 
}
\author{Dawn Lawrie,$^\dagger$ Sean MacAvaney,$^\ddagger$ James Mayfield,$^\dagger$ \\ 
Paul McNamee,$^\dagger$ Douglas W. Oard,${^o}^\dagger$ Luca Soldaini,$^*$ Eugene Yang$^\dagger$}
\affiliation{
  \institution{$^\dagger$Johns Hopkins University Human Language Technology Center of Excellence,\\
  $^\ddagger$University of Glasgow, $^o$University of Maryland, $^*$Allen Institute for AI}
  \country{}
}

\email{lawrie@jhu.edu,sean.macavaney@glasgow.ac.uk,mayfield@jhu.edu}
\email{mcnamee@jhu.edu, lucas@allenai.org, oard@umd.edu, eugene.yang@jhu.edu}

\renewcommand{\shortauthors}{Lawrie et al.}

\begin{abstract}

The principal goal of the TREC Neural Cross-Language Information Retrieval (NeuCLIR) track
is to study the effect of neural approaches on cross-language information access.
The track has created test collections containing Chinese, Persian, and Russian news stories and Chinese academic abstracts.
NeuCLIR includes four task types: Cross-Language Information Retrieval (CLIR) from news,
Multilingual Information Retrieval (MLIR) from news,
Report Generation from news,
and CLIR from technical documents.
A total of 274 runs were submitted by five participating teams
(and as baselines by the track coordinators)
for eight tasks across these four task types.
Task descriptions and the available results are presented.
\end{abstract}

\settopmatter{printfolios=true}
\maketitle

\section{Introduction}

This is the third and final year of the TREC Neural Cross-Language Information Retrieval (NeuCLIR) track.\footnote{\url{https://neuclir.github.io}}
The first year of the track included News CLIR tasks~\cite{lawrie2022overview}.
The second year of the task added News MLIR and Technical Documents CLIR tasks~\cite{lawrie2023overview}.
In this third and final year of NeuCLIR those three task types continued,
and we added a new Report Generation pilot task.
This overview describes NeuCLIR's each of these four task types.

There are three News CLIR tasks,
each of which has topics in English and news documents in one other language
(Chinese, Persian or Russian).\footnote{The news test collections also provide topics in Chinese, Persian and Russian, which can be used to create monolingual baselines or for other purposes.}
CLIR is the most mature of the track's task types,
and the capabilities that CLIR provides are foundational to the other three task types.
Current CLIR systems face two broad challenges that distinguish CLIR from monolingual retrieval:
(1) there is less robust training data than is available for monolingual ranked retrieval tasks; and 
(2) there is misalignment of term representations for different languages in multilingual embeddings.
The news test collections in Chinese, Persian and Russian are the same as in the TREC 2022 NeuCLIR track.
In 2024, the CLIR tasks provide relevance judgments for 56 new topics in Chinese,
68 new topics in Persian, and  65 new topics in Russian.
Three of the five participating teams submitted CLIR runs in 2024, and the track coordinators also created baseline runs.

There is one News MLIR task,
in which the topics are in English and the documents to be searched
are news stories from the union of the Chinese, Russian and Persian news collections.
This task requires generating a single ranked list for each topic over the unified collection.
The principal additional challenge in this task is that scores computed independently for documents
in different languages might not be as easy to compare
as are scores for documents in a single language.
This is the second year of the News MLIR task.
In 2024, the news MLIR task has relevance judgments for 51 topics,
each of which is also present in the 2024 topic set for two or more of the news CLIR tasks. 
Four of the five participating teams submitted results for the news MLIR task in 2024,
and the track coordinators also created baseline runs.

There is one Technical Documents CLIR task
in which the topics are in English and the documents to be searched are Chinese academic abstracts.
The principal additional challenge in this task is that some standard tools
such as multilingual embeddings, pretrained language models such as BERT~\cite{DBLP:conf/naacl/DevlinCLT19},
or generative large language models
may be less well suited to the highly-technical language of these abstracts than to more general news text.
In 2024, the Technical Documents CLIR task has relevance judgments for 71 new topics.
Three of the five participating teams submitted results for the Technical Documents CLIR task,
and the track coordinators also created baseline runs.

Report Generation is a new task type this year. 
In the Report Generation task,
systems receive a report request in English
and are asked to respond with an English report that meets the requirements specified in the request,
and in which substantive statements in the report are supported by a citation to one or more documents
in one of the three news collections.
There are three report generation tasks,
one each for reports with citations to Chinese documents, Russian documents, and Persian documents.
Evaluation used a recently-proposed evaluation framework for such reports~\cite{mayfield2024evaluation}.
Four of the five participating teams submitted results for the Report Generation task.

The remainder of this paper is organized as follows.
We begin with a brief summary of the News CLIR and News MLIR tasks,
emphasizing changes since TREC 2023.  
This is followed by results for those tasks.
Next, we present a brief summary of the Technical Documents CLIR task and the results for that task.
Additional details for the News CLIR tasks can be found
in the TREC 2022 and TREC 2023 NeuCLIR track overview papers~\cite{lawrie2022overview,lawrie2023overview};
additional details on the News MLIR and Technical Documents CLIR tasks can be found in the TREC 2023 track overview paper.
Following that, we provide details for our new Report Generation task. 
This will be the final year of the TREC NeuCLIR track.
Report generation will continue to be evaluated in a new TREC 2025 RAGTIME track,
so we conclude with a brief overview of our plans for RAGTIME. %

\section{News Retrieval Tasks}

In this section we describe the \clirtask\ and MLIR tasks.

\subsection{Task Definitions}

We have two news retrieval task types, \clirtask\ and MLIR.
The News CLIR task includes two ``settings'' (i.e., sub-tasks):
ad hoc CLIR  and (for pool enrichment) monolingual retrieval.
The two settings use the same document collections, topics, and relevance assessments.
Monolingual runs use topics manually translated into the language of the documents;
ad hoc runs use the original English topics and rank documents from the entire collection.

\subsubsection{Ad Hoc CLIR}
\label{subsub:clir-ad-hoc}
For the ad hoc CLIR setting,
systems receive a document collection in Chinese, Persian, or Russian,
and a set of topics written in English.
For each topic, the system must return a ranked list of 1,000 documents
drawn from the entire target language document collection,
ordered by likelihood and degree of relevance to the topic.
Runs that use a human in the loop for ad hoc retrieval
(or had design decisions influenced by human review of the topics)
are indicated as ``manual'' runs;
all others are considered ``automatic.''

\subsubsection{Monolingual Retrieval}%
While monolingual retrieval is not a focus of the NeuCLIR track,
monolingual runs can improve assessment pools
and serve as points of reference for cross-language runs.
The monolingual retrieval setting is identical to the ad-hoc setting,
but it uses topic files that are human translations of the English topics
into a target language in a way that would be expressed by native speakers of the language.
This setting is suitable for teams looking to explore monolingual ranking in languages other than English.
It also has a lower barrier to entry than do the cross-language tasks.

\subsubsection{Multilingual Information Retrieval (MLIR) Task}

NeuCLIR 2023 added a multilingual retrieval task. 
This task is identical to the \clirtask\ task described in \S\ref{subsub:clir-ad-hoc},
except systems must search all three document collections and produce a single unified ranked list. 
In other words, systems should treat the three document collections (\S\ref{sub:documents})
across all three languages as a single corpus. 
Participants were informed that, since topics for this task are derived from those of the \clirtask\ task,
there is no guarantee that each topic will have relevant results in every language.
\subsection{Documents}
\label{sub:documents}

NeuCLIR 2024 continues to use the NeuCLIR~1 document set,
which was also used for NeuCLIR 2022 and 2023.
The collection consists of roughly two million Persian documents,
three million Chinese documents,
and almost five million Russian documents
spanning the years 2016 to 2021.
For more information about how to extract text from Common Crawl News documents
and how the collection can be obtained, see the NeuCLIR 2022 Overview paper~\cite{lawrie2022overview}. 

This year we discovered a problem with language identification
that affected 1,157 CommonCrawl files of the 16,951 files in the time window of August 2016 to July 2021.
None of the documents in the affected files is included in the collection,
which means that after pre-filtering and de-duplication,
765,299 Chinese documents, 317,392 Persian documents, and 3,410,884 Russian documents were excluded from the collection.
This has less impact on the Russian collection, which was down-sampled to five million documents.
For the other two languages, the collections would likely have included about half of the missing documents.  %
\begin{table*}[]
\caption{Relevance judgment statistics for 2024 News topics. }\label{tab:judgment}
    \centering
\begin{tabular}{l|ccc|c}
\toprule
Topics Developed & Chinese & Persian & Russian & MLIR \\
\midrule
\# Topics Retained          &   56    &  68     &  64      & 52 \\
Avg. \# Judgments / Topic &   700    &   661    &   577     & 1999 \\
Avg. Prevalence           &    0.022    &  0.036     &    0.040    & 0.022 \\
\bottomrule
\end{tabular}

\end{table*}

\subsection{Topics}
\label{sec:topics}

NeuCLIR 2024 topic development was completed entirely by NIST assessors. 

The topic development process was identical to that used in 2023.
Two paired assessors with language skills in two different languages
met virtually to brainstorm a topic together.
Good topics were described as topics that ``revolve around events, people, and places,
and [are] %
significant enough to have coverage in more than one language.''
After exploring the collection with monolingual searches in the two assessor languages,
a description was written,
followed by a first draft of the narrative,
and finally the title.
This year assessors used a neural retrieval engine in hopes that initial statistics 
would better capture the prevalence of a topic in the collection.
As usual a single %
monolingual search was initiated in each language,
and each assessor counted the number of relevant documents in the top twenty-five returned documents.
They were instructed to revise the topic if the count in either language was less than one or greater than twenty.
Once the topic was appropriately scoped,
the narrative was revised and the topic was included in the topic set for 2024.
Ninety-two topics were created.

Because it was the third year using this document set,
some topics were removed from consideration
because they were semantically too similar to topics that appeared in 2022 or 2023.
In addition,
some topics were released as development data for the report generation task.
Since the report generation development data included associated documents,
these topics were also not considered for inclusion in the topic set.
Each language ended up with a different number of assessed topics
because the languages differ in the time required to judge a topic.
\subsection{Relevance Judgments}\label{sec:clir:rel-judgment}

Once all submissions were made,
by-language pools were created that integrated the top-ranked documents from both \clirtask\ and MLIR runs
as well as documents cited by reports for the associated report request if the topic had one.
The top 100 documents for runs that teams prioritized as their top nine runs were included in the pools.
Such runs have a checkmark in the JFD columns of Tables~\ref{tab:zho-full-results}, \ref{tab:fas-full-results}, and \ref{tab:rus-full-results}. 
For other runs, a depth of fifty was used. 
The Coordinators created some runs based on last year's system submissions as baselines.
The Coordinators were considered to be a unique team and followed the same cutoffs as other teams.
The same four-point scale as NeuCLIR22~\cite{lawrie2022overview} was used to judge relevance.
The four-point scale was converted to a three-point scale for the qrels,\footnote{The mapping from four relevance grades was 3->3, 2->1, 1->0, and 0->0. }
again as was done for NeuCLIR22~\cite{lawrie2022overview} and NeuCLIR23~\cite{lawrie2023overview}.

After pooling, some topics were dropped from the \clirtask\ and MLIR tasks according to the following rules: 
\begin{itemize}
\item
If more than 40\% of the judged documents were judged to be somewhat or very valuable 
in a particular language,
drop the topic from the \clirtask\ task in that language and from the MLIR task. 
\item
If the relevance judgments for a topic had fewer than two documents in the somewhat or very valuable categories in a language,
drop that topic from the \clirtask\ task in that language,
but include it in the MLIR task. %
\item If a topic had fewer than two relevant documents across all languages, drop it from the MLIR task.
\end{itemize}
Of the 92 topics created, 56 were retained for Chinese, 68 for Persian, 64 for Russian, and 52 for the MLIR task. 
All MLIR topics are judged in all three languages, though not all topics have relevant documents in all languages. 
Table~\ref{tab:judgment} describes features of the topics that were retained
based on the criteria described in Section~\ref{sec:topics}.

\subsection{Additional Resources}
\label{sec:clir-resources}
To support the aims of the \clirtask\ and MLIR tasks
and to lower the barrier to entry for new participants,
the track made available additional resources beyond the document collection and topics.
These resources included machine translated versions of queries and document collections,
translations of the widely used MS~MARCO collection into the three NeuCLIR-1 document languages,
and previously-used IR test sets in the three NeuCLIR languages.

Machine translated versions of the queries were created by the online \textit{Google Translate} service.

English versions of the document collections were created by directional machine translation Transformer models
using the Sockeye version~2 toolkit.
As the document collection for the \clirtask\ and MLIR tasks did not change from 2022,
these are the same translations that were provided in the previous two years of the track
and that are described in the TREC 2022 NeuCLIR Overview paper\cite{lawrie2022overview}.

As many neural IR systems are trained using data derived from the MS~MARCO dataset\cite{bajaj2018ms},
translations of this English resource into different languages were provided.
We provided a version on Hugging Face called \textit{NeuMARCO}.\footnote{\url{https://huggingface.co/datasets/neuclir/neumarco}}
We also provided links to similar translations from the \textit{mMARCO} project\cite{bonifacio2022mmarco} on the NeuCLIR website.

The track website also collected a number of multilingual and bilingual resources in the languages of the track
including HC4 -- a CLIR collection built over three years of Common Crawl data in the same three languages~\cite{hc4},
as well as two multilingual CLIR datasets based on Wikipedia,
known as CLIRMatrix~\cite{clirmatrix} and WikiCLIR~\cite{wikiclir}.

Finally, the topics and relevance judgments from the previous iterations of NeuCLIR (2022 and 2023)
were available to track participants,
either from NIST, or in \texttt{ir\_datasets}.\footnote{\url{https://ir-datasets.com}}
These datasets could be used for system tuning and validation.
\begin{table*}[]
\caption{Number of News CLIR and MLIR runs submitted by language and team. Incorrectly submitted runs by \texttt{h2oloo} are reassigned to the correct tasks in the table but not in the actual pools created for relevance judgments. }\label{tab:clir-participation}
    \centering
    
\begin{tabular}{l|cccc|c}
\toprule
Team & Persian & Russian & Chinese & MLIR & Total \\
\midrule
IRLabAmsterdam & 0 & 0 & 0 & 3 & 3 \\
ISI-SEARCHER & 1 & 1 & 1 & 1 & 4 \\
h2oloo$^*$ & 11 & 11 & 11 & 5 & 38 \\
hltcoe & 15 & 15 & 15 & 12 & 57 \\
Track Coordinator Baselines & 21 & 21 & 21 & 8 & 71 \\
\midrule
Total & 37 & 37 & 37 & 62 & 173 \\
\bottomrule
\end{tabular}
\end{table*}

\subsection{Participation}

The News MLIR task had four participating teams, three of which also submitted runs to News CLIR tasks:
\begin{itemize}
    \item Johns Hopkins University HLTCOE~\cite{participants-hltcoe}
    \item University of Amsterdam (MLIR only)~\cite{neuclir-participants-ams}
    \item University of Southern California ISI~\cite{neuclir-participants-isi}
    \item University of Waterloo~\cite{neuclir-participants-h2oloo}
\end{itemize}
In addition to track participants,
the track coordinators contributed baseline runs to ensure representation of
a wide variety of retrieval approaches in the judgment pools.
Table~\ref{tab:clir-participation} shows the number of runs submitted under each category.

\subsection{Track Coordinator Baselines}
\label{coord_runs}

The track coordinators also prepared several runs to include as baselines.
The foci of these runs were monolingual dense retrieval and sparse retrieval.
The run names are outlined in Table~\ref{tab:run-name}.
Notice that the MLIR run conditions are a subset of the CLIR run conditions.
This is because there are fewer options for sparse retrieval when the documents are in multiple languages,
and the idea of monolingual retrieval is nonsensical in this scenario.

\subsubsection{Monolingual Dense Retrieval}

For monolingual non-English retrieval, ColBERT models were trained using translate-distill~\cite{tdistill}.
While the student model was shown translated queries and documents from MS~MARCO in the non-English setting
(e.g., Chinese for Chinese retrieval),
scores came from the teacher model applying scores to the English queries and documents.
This way the scores were not influenced by any `translationese' introduced by machine translation. 
These runs used the human-translated queries along with the native documents.
Because there is no notion of monolingual retrieval in MLIR, this approach is not used in MLIR.

\subsubsection{Sparse Retrieval}

The coordinators included two broad categories of sparse retrieval.
Probabilistic Structured Queries relies on a probabilistic translation table to cross the language barrier.
BM25 relies on translation to cross the language barrier.
Machine translation of either the query or document side each resulted in a CLIR run.
Monolingual runs rely on queries expressed in the document language.
Since NeuCLIR produces queries first in English,
these non-English queries are referred to as human translations.

Probabilistic Structured Queries (PSQ)~\cite{darwish2003probabilistic}
is a translation approach that probabilistically matches a token from one language
to a distribution of tokens in another. 
This technique can be used to translate queries, documents, or both. 
Prior work~\cite{wang2012matching} has concluded that mapping documents to the query language at indexing time
achieves the best effectiveness while minimizing query latency. 
The resulting documents are bags of probabilistic tokens in the query language. 
They can be indexed as ordinary documents in a sparse retrieval model such as BM25 or HMM with real-valued weights. 

Our submission uses PSQ~\cite{yang2024efficiency} to translate the documents
and uses a Hidden Markov Model (HMM)~\cite{xu2000cross} for retrieval.

The Patapsco framework~\cite{patapsco}
supports CLIR lexical retrieval through Pyserini~\cite{pyserini}.
Patapsco ensures that language-specific processing is consistent for both queries and documents.
The coordinators submitted BM25 monolingual runs that used human-translated queries
to search documents in their native language (QHT),
CLIR runs that used the track-provided machine query translations to search the native documents (QGT),
and runs that used English queries to search the track-provided document translations (DT).
All languages used spaCy~\cite{spacy} for tokenization.
For Russian and English machine translation, spaCy also provided stemming,
while Parsivar~\cite{Mohtaj2018ParsivarAL} was used for Persian stemming.
(We did not stem Chinese.)
We explored three query variants: title, description, and title+description. 
We also explored the addition of ten RM3 expansion terms.

\begin{table*}[]
\caption{Track Coordinator baseline runs for CLIR and MLIR tasks. Run names in italics are monolingual runs. }\label{tab:run-name}
    \centering
    
\begin{tabular}{l|cl|c|l}
\toprule
Run Name & Type &  Model & Query & Description \\
\midrule
\multicolumn{4}{l}{News CLIR and Technical Documents CLIR Baseline Runs}  \\
\midrule
\it{patapscoBM25htRM3desc}                 & Sparse & BM25 & D & Monolingual Patapsco with RM3 \\
\it{patapscoBM25htRM3td}               &  Sparse &    BM25 & TD & Monolingual Patapsco with RM3 \\
\it{patapscoBM25htRM3t}                         & Sparse &  BM25 & T & Monolingual Patapsco with RM3 \\
\it{patapscoBM25htnoRM3desc}             &  Sparse &    BM25 & D & Monolingual Patapsco without RM3 \\
\it{patapscoBM25htnoRM3td}                 &  Sparse &  BM25 & TD & Monolingual Patapsco without RM3 \\   
\it{patapscoBM25htnoRM3title}           &  Sparse &    BM25 & T & Monolingual Patapsco without RM3 \\   
patapscoBM25qtRM3desc                 & Sparse & BM25 & D &  Patapsco Google query translation with RM3 \\
patapscoBM25qtRM3td               &  Sparse &    BM25 & TD & Patapsco Google query translation with RM3 \\
patapscoBM25qtRM3title                         & Sparse &  BM25 & T & Patapsco Google query translation with RM3 \\
patapscoBM25qtnoRM3desc             &  Sparse &    BM25 & D & Patapsco Google query translation without RM3 \\
patapscoBM25qtnoRM3td                 &  Sparse &  BM25 & TD & Patapsco Google query translation without RM3 \\   
patapscoBM25qtnoRM3title           &  Sparse &    BM25 & T & Patapsco Google query translation without RM3 \\   
patapscoBM25dtRM3desc                 & Sparse & BM25 & D &  Patapsco indexing translated documents with RM3 \\
patapscoBM25dtRM3td               &  Sparse &    BM25 & TD & Patapsco indexing translated documents with RM3 \\
patapscoBM25dtRM3title                         & Sparse &  BM25 & T & Patapsco indexing translated documents with RM3 \\
patapscoBM25dtnoRM3desc             &  Sparse &    BM25 & D & Patapsco indexing translated documents without RM3 \\
patapscoBM25dtnoRM3td                 &  Sparse &  BM25 & TD & Patapsco indexing translated documents without RM3 \\   
patapscoBM25dtnoRM3title           &  Sparse &    BM25 & T & Patapsco indexing translated documents without RM3 \\   
\it{plaid\_distill\_mono\_ht} &  Dense &        PLAID & TD & TD PLAID with monolingual training  \\
fast\_psqtd                               & Sparse &          PSQ & TD & PSQ-HMM \\
fast\_psqtitle                                & Sparse &          PSQ &  T & PSQ-HMM \\
\midrule
& \multicolumn{4}{l}{News MLIR Baseline Runs}  \\
\midrule
patapscoBM25dtRM3desc                &  Sparse &    BM25 & D & Patapsco indexing translated documents with RM3 \\
patapscoBM25dtRM3td                &  Sparse &    BM25 & TD & Patapsco indexing translated documents with RM3 \\
patapscoBM25dtRM3title                &  Sparse &    BM25 & T & Patapsco indexing translated documents with RM3 \\
patapscoBM25dtnoRM3desc                &  Sparse &    BM25 & D & Patapsco indexing translated documents without RM3 \\
patapscoBM25dtnoRM3td                &  Sparse &    BM25 & TD & Patapsco indexing translated documents without RM3 \\
patapscoBM25dtnoRM3title                &  Sparse &    BM25 & T & Patapsco indexing translated documents without RM3 \\
fast\_psqtd                            & Sparse &          PSQ & TD & Combining CLIR PSQ-HMM scores using score fusion\\
fast\_psqtitle                             & Sparse &          PSQ &  T & Combining CLIR PSQ-HMM scores using score fusion\\
\bottomrule
\end{tabular}

\end{table*}

\subsection{Results}

\subsubsection{Effectiveness and Run Diversity}

\begin{figure*}[hbtp]
    \centering
    \includegraphics[width=\linewidth]{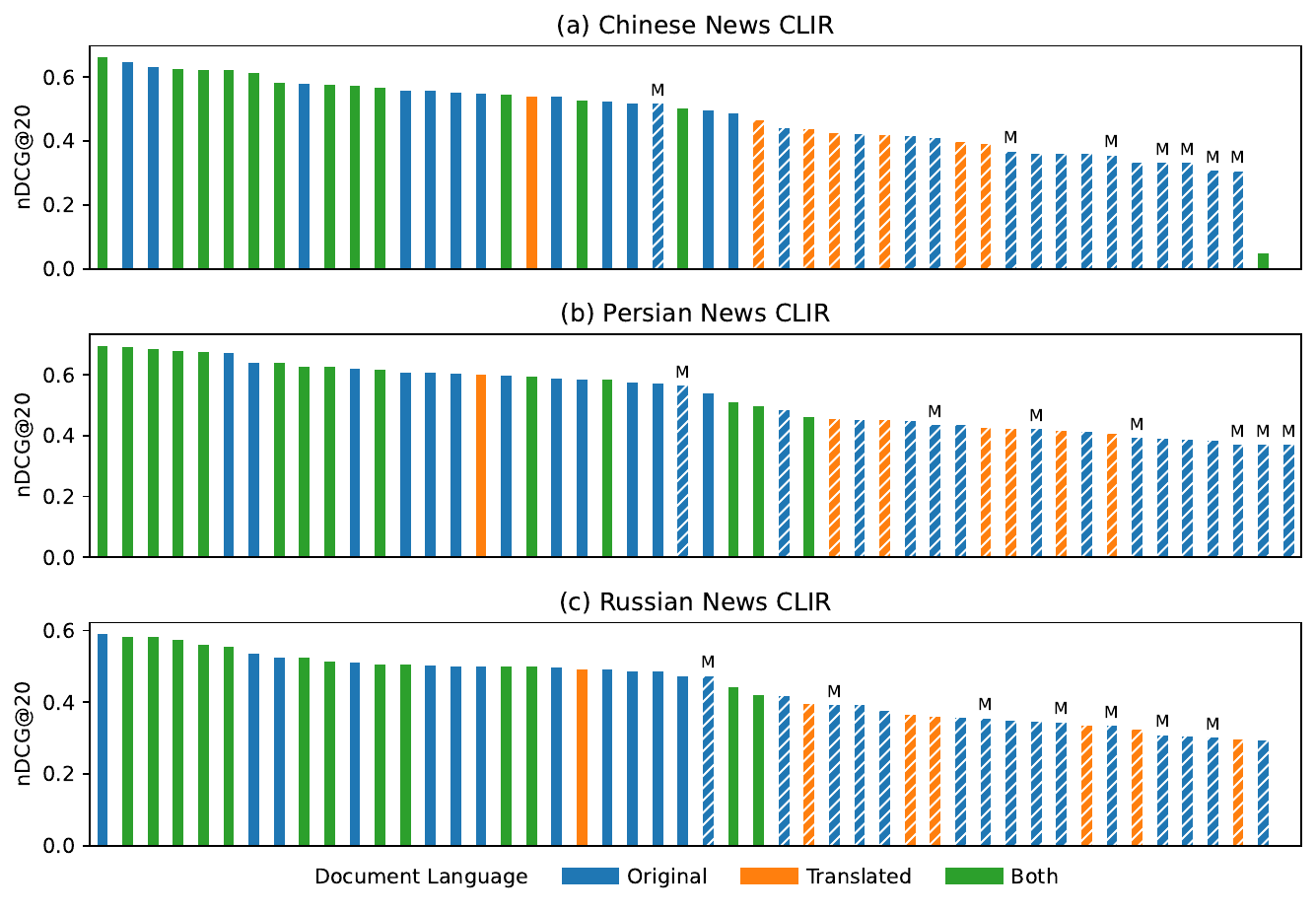}
    \caption{News CLIR nDCG@20. Coordinator runs are marked with slashes. Monolingual runs (i.e., using human-translated topics) are marked with ``M'' at the top of the bar.}\label{fig:clir-ndcg-bar}
\end{figure*}

\begin{figure*}[htp]
    \centering
    \includegraphics[width=\linewidth]{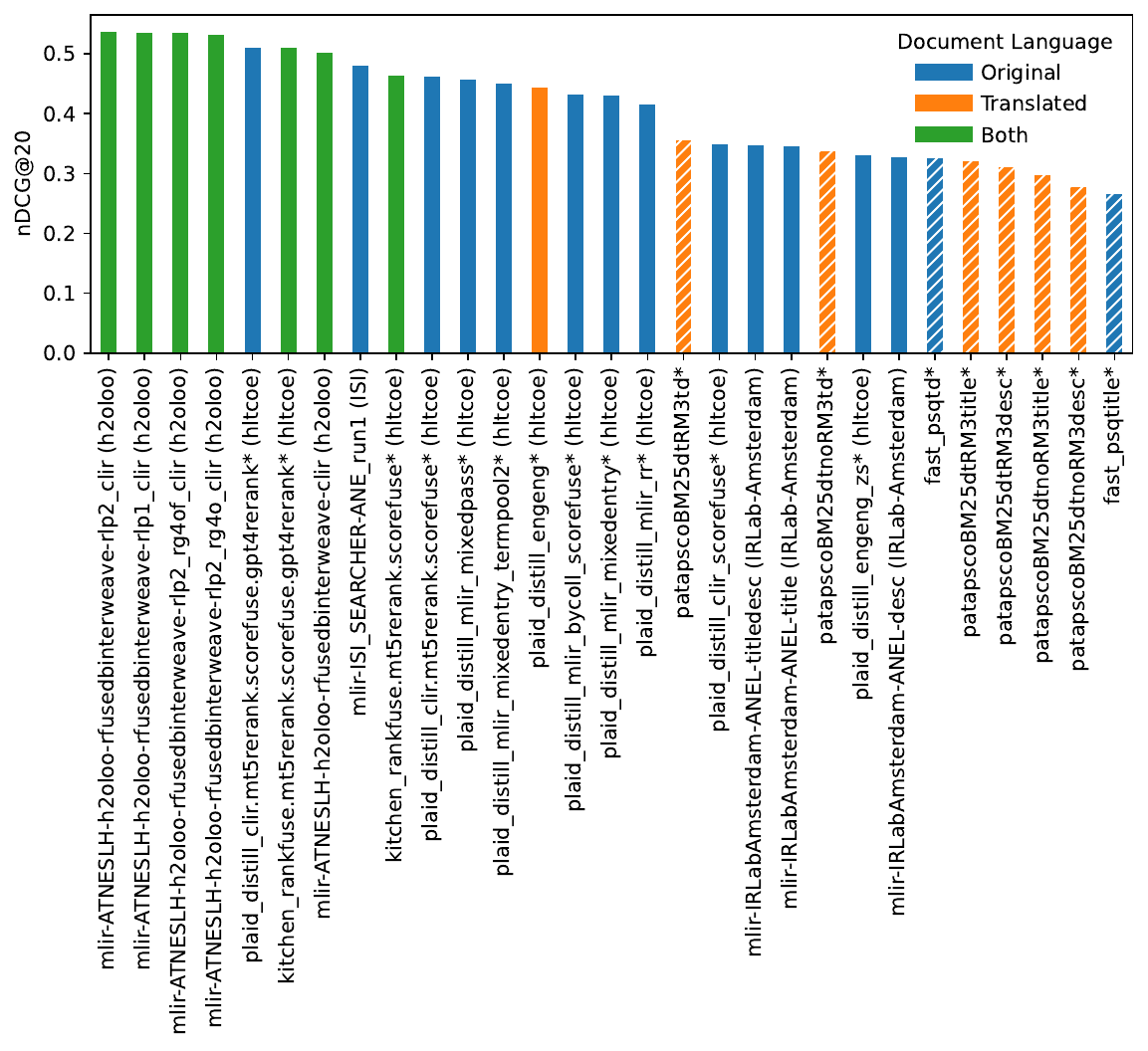}
    \caption{News MLIR nDCG@20. Coordinator runs are marked with slashes. }\label{fig:mlir-ndcg-bar}
\end{figure*}

Submissions this year feature LLM reranking. 
Tables~\ref{tab:zho-full-results}, \ref{tab:fas-full-results},
and \ref{tab:rus-full-results} present the evaluation results of the News CLIR submissions. 
Based on their run ids, all top-scoring runs use a generative model as the final-stage reranker,
which has been the trend over the past year in the retrieval research community. 
Interestingly, as summarized in Figure~\ref{fig:clir-ndcg-bar}
the most effective system for the Russian task used only the original documents,
unlike others that used both original and translated documents. 
Such systems show the potential of developing effective neural retrieval pipelines
without using any machine translation during indexing and search. 
While this year marks the final year of NeuCLIR,
we believe there is still great room for future improvement in CLIR tasks. 

The News MLIR task received fewer submissions but still included a wide range of approaches. 
nDCG@20 results are summarized in Figure~\ref{fig:mlir-ndcg-bar}.
Please refer to Table~\ref{tab:mlir-full-results} for the full results. 
One noticeable difference from the CLIR tasks is the emphasis on fusion techniques. 
Top-scoring runs in the MLIR task use systems that fuse multiple runs,
possibly including CLIR runs, before reranking.
Such approaches exemplify the need for strong end-to-end MLIR first (or early) stage retrievers
that provide strong coverage in all languages. 
Similar to the CLIR tasks, top-scoring runs all use both original and translated documents. 

All participant submissions include neural models in their pipelines, a common theme in modern retrieval research. 
To ensure that the pools still include documents that have surface forms matching with the queries,
the coordinators submitted a number of BM25 variations to enrich the pools. 
While most of these runs are less effective than participant submissions,
they ensure the quality of our pools and of the resulting collection. 

\subsubsection{Reusability}

\begin{figure*}
    \centering
    \includegraphics[width=\linewidth]{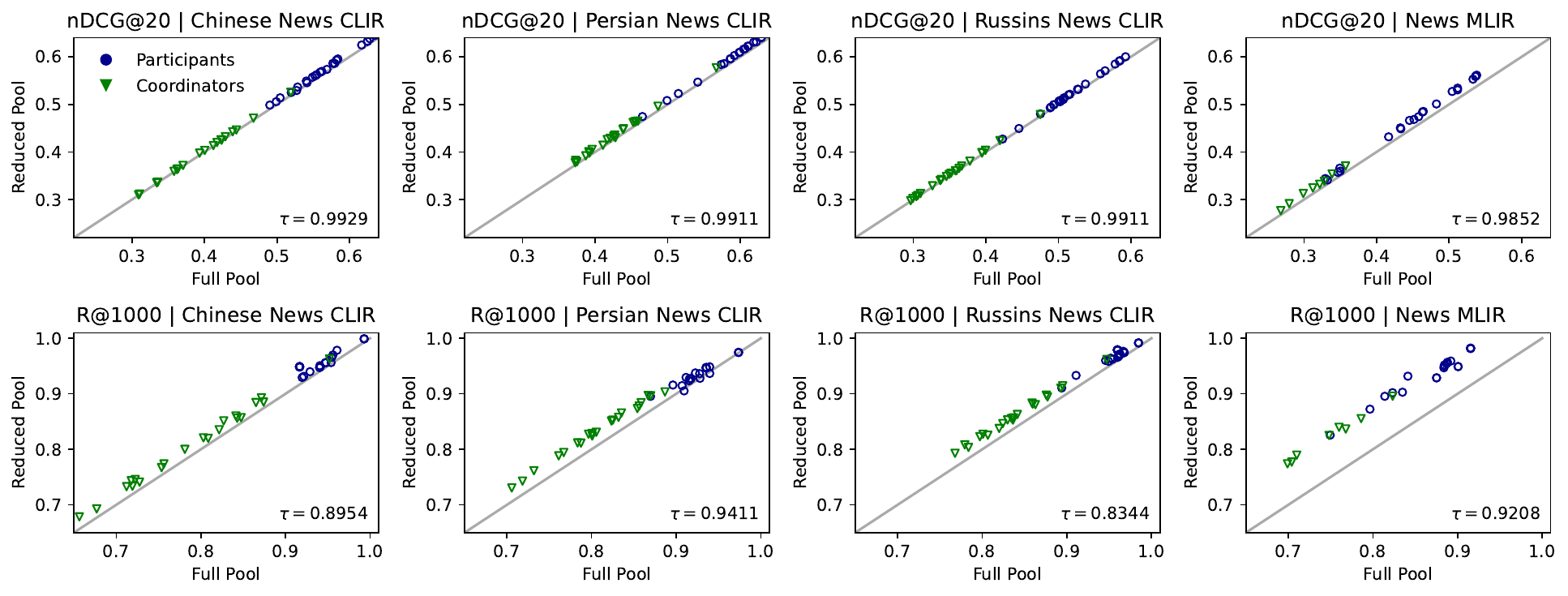}
    \caption{News CLIR and MLIR Leave-One-Run-Out Experiments.}\label{fig:news-loo}
\end{figure*}

\begin{figure*}
    \centering
    \includegraphics[width=\linewidth]{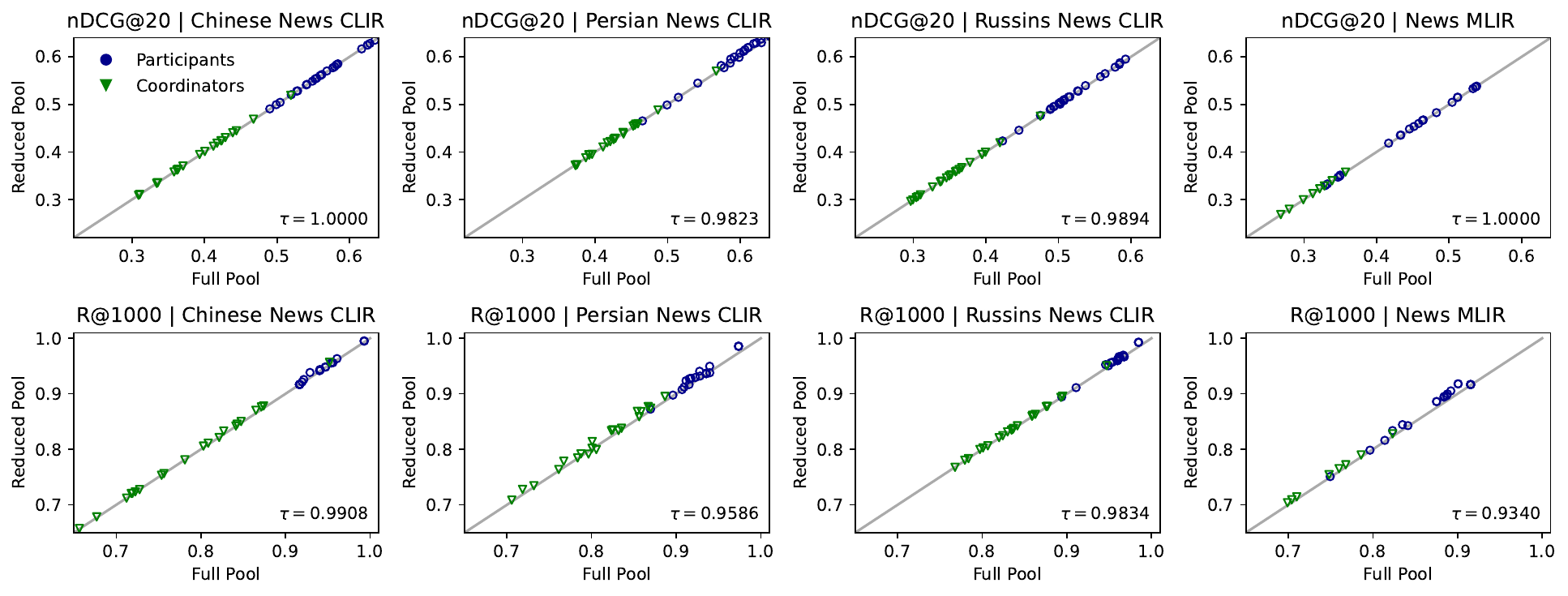}
    \caption{News CLIR and MLIR Leave-One-Team-Out Experiments.}\label{fig:news-loto}
\end{figure*}

We conducted leave-one-run-out and leave-one-team-out experiments to assess the reusability of the collection. 
Since the pools for the CLIR and MLIR tasks are constructed from both CLIR and MLIR runs,
leave-one-team-out removes all runs from the team across both CLIR and MLIR tasks to simulate the absence of the team entirely. 
Figures~\ref{fig:news-loo} and \ref{fig:news-loto} summarize the results of these experiments.

There are many reranking runs,
so the differences between the top-ranked document sets across runs are small
(though the ranking may be drastically different, resulting in different effectiveness); 
this is shown in Figures~\ref{fig:zho-overlap}, \ref{fig:fas-overlap}, \ref{fig:rus-overlap}, and \ref{fig:mlir-overlap}. 
nDCG@20 is stable when leaving runs out of pooling across both CLIR and MLIR tasks. 
However, the relevant documents brought in by each run are more different,
resulting in larger differences in R@1000. 
When leaving a team out from pooling, the reduced pools provide similar metric values
and thus stable system ordering. 

It is rare to see leave-one-run-out lead to less stable pools than leave-one-team-out experiments. 
We suspect this phenomenon is due to the submission error of one of the participating teams,
who submitted all their MLIR runs to the CLIR task and vice versa; 
this resulted in shallower pool depths for their runs. 
In this experiment, we reassigned each run to its correct task;
however, the actual pools were already created using the errorful runs. 
Despite this incident, we believe the collection to be reusable. 

\section{Technical Documents Task}

The Technical Documents Task is in its second year.
It is a cross-language ad hoc retrieval task, with English queries and Chinese documents.
The key distinguishing feature of this task is the technical nature of the documents.
The document collection is the same as was used for the Technical Documents Task in 2023.
While last year's pilot task used a small number of topics,
this year's full task allows researchers to gauge the effectiveness of existing CLIR approaches on technical documents,
and to identify along which dimensions those systems need improvement.

This task contains the same two settings as the newswire \clirtask\ task,
namely ad hoc CLIR, and monolingual retrieval.
\subsection{Documents}

The documents for this task were abstracts from the Chinese Scientific Literature (CSL) dataset~\cite{li-etal-2022-csl}.
The dataset contains 396,209 journal abstracts from 1,980 academic Chinese journals spanning 67 general disciplines,
where Engineering, Science, Agriculture, and Medicine dominate.
This is the same document set as was used in NeuCLIR 2023 for the technical documents task~\cite{lawrie2023overview}. %
\subsection{Topics}

Topic creation in 2024 was accomplished by fifteen graduate students and one postdoc
in Biology, Computer Science, Earth Science, Economics, Engineering, Math, and Physics %
from The Johns Hopkins University and from the University of Maryland, College Park.
Annotators were hired based on their Chinese language skills and their familiarity with scientific research.
During an interview, students were asked to describe their research area in both Chinese and English.
They were then asked to choose a research topic they were familiar with,
enter an English, Chinese, or mixed language query on that topic into an interactive search system
that returned documents from the CSL dataset,
and read and briefly summarize the top returned documents
to determine whether they were relevant to their search.
The purpose of this part of the interview
was to ensure that the collection contained documents related to their area of research
and to determine whether they could assess documents accurately in a timely fashion.
Of the fifteen students, eleven were Ph.D. students and the others were Masters students.

Once hired, each annotator participated in a three-hour online training session.
During the training, the topic creation task was explained.
Then each person worked independently to create their first topic.
During that process, two of the Coordinators reviewed their ongoing work.
This exercise was used to ensure that topics had a suitable level of specificity,
and that the tool was being used properly
to determine whether abstracts on the topic existed in the collection.
After the training, assessors were asked to spend up to a total of ten hours creating five to eight topics.
One assessor created five topics, six assessors created six topics apiece, seven assessors created seven topics, and two assessors created eight topics,
yielding a total of 106 topics.

The English %
title, description, and narratives were reviewed by a coordinator
to ensure that the topic was sufficiently descriptive.
The topic %
was also checked for grammar and spelling.
In some cases, the assessors were asked to revise topics that appeared to be too vague or were not understandable. 
Narratives were also given special attention to ensure they were sufficiently detailed.
After any revision, assessors checked the translation to ensure that it incorporated any changes.
The translations did not undergo any external quality control.
In the end 106 topics were distributed to participants. %
\subsection{Relevance Judgments}

\begin{table*}[thp]
\caption{Relevance judgments for the Technical Documents CLIR task.}\label{tab:tech-judge}
    \centering

\begin{tabular}{l|cccccc|c}
\toprule
                                  &  Biology &  Computer Science &  Earth Science &  Economics &  Engineering & Physics &  Overall \\
\midrule
\# Topics Included                  &          13 &     14 & 19 &  2 &  10 & 13 & 71 \\
\midrule
Avg. \# Judgments / Topic         &     367.92 &     470.43 & 292.53 & 375 & 315.2 & 311.54 & 350.41  \\
\midrule
Avg. \# Somewhat Valuable / Topic &       8.69 & 17.36 & 10.58 & 28 & 12 & 7.08 & 11.62\\
Avg. \#  Very Valuable / Topic    &       4.46 & 9.07 & 9.11 & 27.5 & 9.8 & 6.46 & 8.46 \\
\bottomrule
\end{tabular}

\end{table*}

Assessors then participated in a second online training session, lasting two hours,
that focused on relevance judgment.
In addition, instructions written in English were provided for completing relevance judgments.
Relevance for the Technical Documents pilot task differed somewhat from the usual TREC view of relevance.
Assessors were asked to imagine that they were writing the background section or the related work section
of a scientific paper on the topic they had created.
They were asked to evaluate whether they would plan to read the paper being judged based on its abstract
so as to possibly cite the paper in their related work section.

They answered two questions about each document:
\begin{definition}
 Does this document contain central information?
 \begin{description}
 \item
 [Yes] There is information in the abstract related to their search topic. 
 \item [No] 
There is no information in the abstract related to their search topic. 
\item [Unable to judge] The document was not viewable in the document viewer panel.
\end{description}
\end{definition}
\begin{definition}

How valuable is the most important information in this document?
 \begin{description}
 \item
[Very Valuable] One would definitely read the paper associated with this abstract when writing the related work section for this research topic.
\item
[Somewhat Valuable] If one had enough time one would read the paper,
because it might have something that could appear in the related work section,
but confidence about that is low.  
\item
[Not that Valuable] One is unlikely to read the paper
because one does not expect to find in it information that one would cite in the related work section. 
\end{description}
\end{definition}

After training, they were asked to judge part of one topic and then given an opportunity to ask questions.
Most assessors finished the first topic the same day as the training.

Assessors were asked to spend at most ten hours judging,
and their progress was tracked.
Some assessors ran out of time, 
leading to seven unjudged topics.
Topics were removed for having fewer than three relevant documents
(affecting eight topics)
or for judging more than 20\% of the pool to be somewhat or very valuable
(affecting sixteen topics).
Four other topics were removed because the assessor experienced technical difficulties while judging the task. 
No assessor had both a topic removed for having too few relevant documents
and a topic removed for having too many relevant documents.

Pools were created from the top thirty-five documents of each submitted run.
In the end seventy-one topics were used to judge system performance.
Table~\ref{tab:tech-judge} contains information on the judgments,
on the average number of very valuable per topic
and the average number of somewhat valuable per topic in the judged pools. 
Documents that the assessor identified as relevant during topic development
were also included in the pools.

\subsection{Additional Resources}

In addition to the document collection itself
and the resources already described in Section~\ref{sec:clir-resources},
the track provided translations into Chinese of the topic fields,
and translations into English of the document texts.
These translations were obtained from the online \textit{Google Translate} service in 2023.
\subsection{Participation}

The Chinese Technical Documents CLIR task had three participating teams that together submitted 28 runs:
\begin{itemize}
    \item Johns Hopkins University HLTCOE~\cite{neuclir-participants-hltcoe}
    \item University of Southern California ISI~\cite{neuclir-participants-isi}
    \item University of Waterloo~\cite{neuclir-participants-h2oloo}
\end{itemize}
\subsection{Track Coordinator Baselines}

The track coordinators also created several runs to include as baselines.
The foci of these runs were monolingual dense retrieval and sparse retrieval.
The run names are outlined in Table~\ref{tab:run-name}.
See Section~\ref{coord_runs} for more information about the algorithms used in these runs.
\subsection{Results}

\begin{figure*}[htp]
    \centering
    \includegraphics[width=\linewidth]{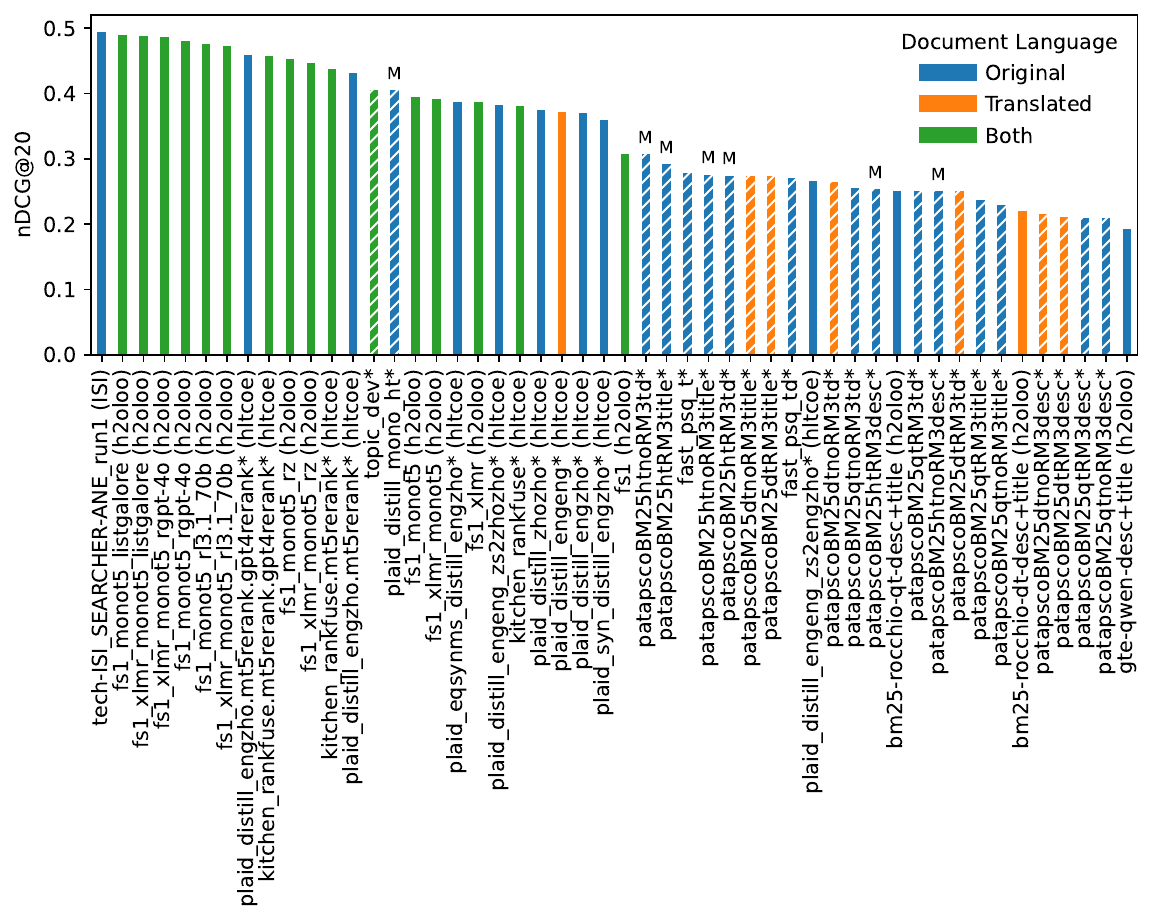}
    \caption{Technical Document CLIR task nDCG@20. Coordinator runs are marked with slashes. Monolingual runs (i.e., using human-translated topics) are marked with ``M'' at the top of the bar. }\label{fig:tech-ndcg-bar}
\end{figure*}

\begin{figure*}[htp]
    \centering
    \includegraphics[width=\linewidth]{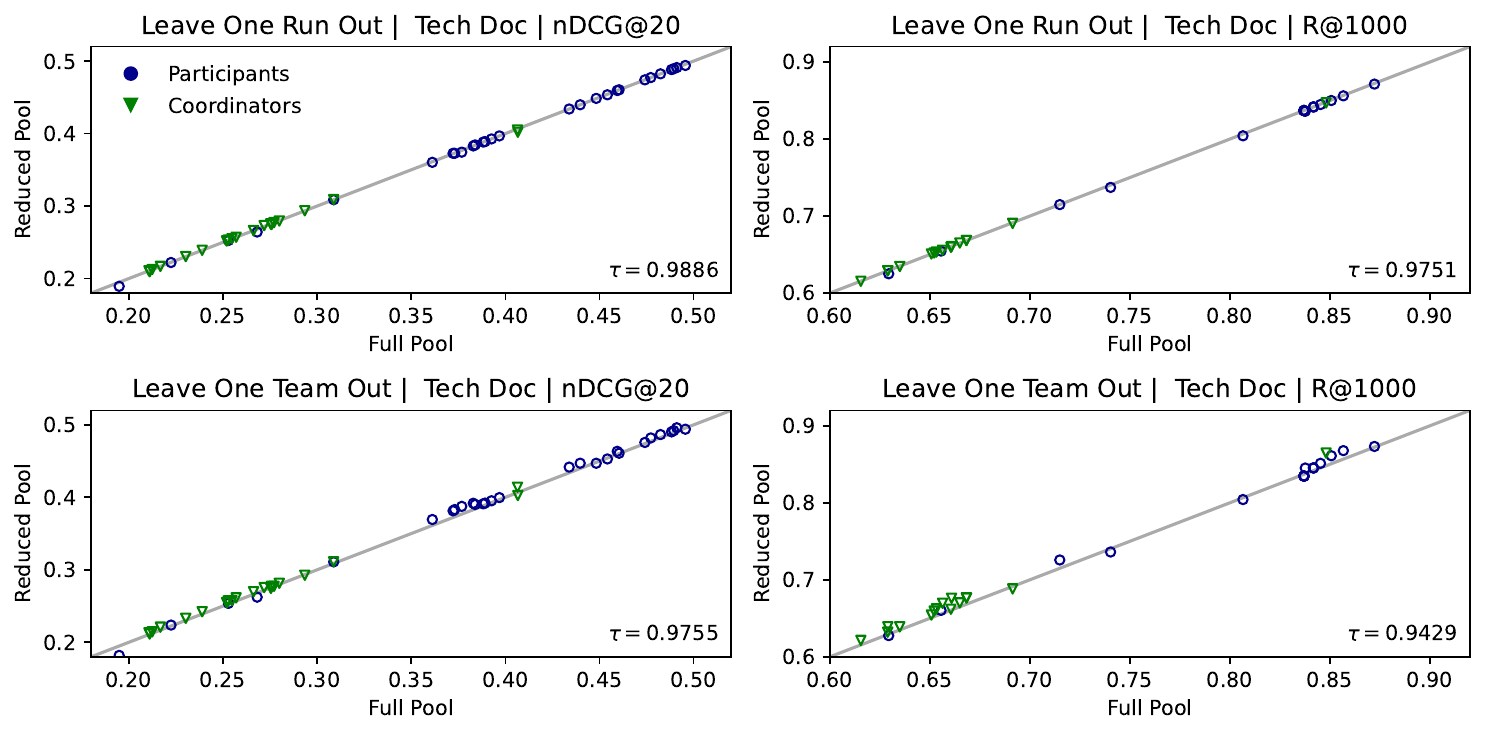}
    \caption{Leave-One-Out and Leave-One-Team-Out Experiments on the Technical Document CLIR task.}\label{fig:loo_exp}
\end{figure*}

\subsubsection{Effectiveness and Run Diversity}

During this final year of NeuCLIR
we primarily focused on the utility and the reusability of the resulting collection. 
Summarized in Figure~\ref{fig:tech-ndcg-bar},
we observe a wide range in effectiveness
with a clear separation between the neural and statistical methods
of around 0.3 in nDCG@20. 
The coordinators submitted the relevant documents found during topic development as a manual run (\texttt{topic\_dev});
among all the runs in Figure~\ref{fig:tech-ndcg-bar}
this run ranked directly below the runs that use heavy rerankers at the end of their retrieval pipeline
Detailed effectiveness measures are presented in Table~\ref{tab:tech-full-results}. 

Most systems used both translated and original document text to increase effectiveness,
with the exception of the best system. 
Top-ranked runs all use a large generative model as a reranker
(based on the run IDs and personal communications with the participants);
this aligns with findings in recent monolingual neural retrieval literature.

Despite the number of participating teams being less than ideal,
the submitted systems included a wide variety of models and approaches.
The coordinators also contributed several statistical runs that are less popular nowadays
and that are missing from the participant submissions. 
While these are significantly less effective than the neural runs,
they provide enrichment to the pools and assurance of measurability on the lower effectiveness ranges. 
Captured in the overlapping graphs in Figure~\ref{fig:tech-overlap},
the coordinator runs are less similar to the participants' runs (darker lower left corner)
for both retrieved documents
(Figure~\ref{fig:tech-overlap}(a))
and retrieved relevant documents
(Figure~\ref{fig:tech-overlap}(b)). 
In particular, the topic development manual run provides a very different set of retrieved documents
(the black strip in Figure~\ref{fig:tech-overlap}(a)) 
than all other system submissions.
Annotators tended to issue several queries.
They tended to stick to one query language,
only occasionally entering mixed-language searches.

\subsubsection{Reusability}

We directly test reusability by conducting leave-one-run-out (LORO)
and leave-one-team-out (LOTO) experiments
to verify the robustness of the set of relevant judgments produced from pooling. 

nDCG@20 on both LORO and LOTO experiments indicates very stable pools
with 0.99 and 0.97 Kendall's $\tau$ respectively.
While leaving one team out from the pool leads to slight instability in the results,
this is due to having only four (including the coordinators) teams in the pool.
However, since teams all provide diverse sets of runs
(especially for \texttt{h2oloo}),
leaving one team out would remove a large and diverse set of runs. 

With the large number of reranking runs, LORO on R@1000 is less meaningful,
since runs from a single organization tend to share retrieval results.
These trends are captured in the overlapping graphs in Figure~\ref{fig:tech-overlap}. 
LOTO indicates a stronger instability in R@1000.
While 0.92 Kendall's $\tau$ is still extremely high,
it indicates less stability than measuring the top of the ranked list, i.e., measuring nDCG@20. 
However, we anticipate no reusability issues for this collection.
Future CLIR research can and should evaluate on this collection. %

\section{Report Generation Pilot Task}

CLIR solves one problem---it ranks documents relative to a query in another language;
but it creates another---someone has to read all those documents!
This is not just a matter of the time and effort required---some searchers may also not be able to read documents in their original language. 
The goal of the TREC 2024 NeuCLIR Report Generation task is to address both of these challenges
by creating concise focused reports
(i.e., multi-document summaries)
in the language of the report request (which in our case is English).
Each report is based on documents from a single NeuCLIR collection (Chinese, Persian or Russian). 
These reports are evaluated based on the degree to which they use correctly cited references
to documents in the specified collection to answer questions that the report requester wished answered
using the procedure proposed by \citet{mayfield2024evaluation}.
\subsection{Documents}

The Report Generation task used the same collections as the News CLIR tasks,
which contain CommonCrawl News articles in Chinese, Russian or Persian.
\subsection{Report Requests}

Assessors began with topics created for the CLIR tasks.
They worked from each of these topics to create a report request
by adding a background section describing why the report was needed
and a detailed problem statement that described what the report should contain.
These added sections were generally based on the topic description and narrative fields of the CLIR topic
from which the report request was derived.
In 2024, relevant documents were assumed to contain answers to the nugget questions
used to evaluate the generated reports;
thus, report requests were intended to ask for the same information as the original topic
as described in that topic's title, description, and narrative fields.

A report request consists of a request ID, a collection ID, a background section, a problem statement,
and a length limit (in Unicode characters).
The background and problem statement fields of a report request are expressed in unstructured text.
Here is an example:
\begin{verbatim}
Background: I am a Hollywood reporter writing an article 
about the highest grossing films Avengers: Endgame and 
Avatar.
Problem statement: The article needs to include when 
each of these films was considered the highest grossing 
film and any manipulations undertaken to bring 
moviegoers back to the box office with the specific 
goal of increasing the money made on the film.
Limit: 2000
\end{verbatim}
\subsection{The Assessment Process}

\begin{table*}[t]
\caption{Report Generation Assessment Statistics. }\label{tab:repgen-assessment-stats}
\centering
\begin{tabular}{l|ccc}
\toprule
{}                             & Chinese & Persian & Russian \\
\midrule
\# of Report Request Developed & 59 & 59 & 59 \\
\midrule
\# of Report Request Assessed  & 21 & 20 & 21 \\
Avg. \# of Nugget Questions per Request    & 13.71 & 13.80 & 13.76 \\
Avg. \# of Nugget Questions w/o Answers in that language per Request &  2.57 & 3.45 & 0.86 \\
\midrule
Avg. \# of Uncaptured but Supported Crucial Nuggets per Assessed Report & 1.57 & 1.89 & 0.55  \\
Avg. \# of Uncaptured but Supported Topical Nuggets per Assessed Report & 1.97 & 3.83 & 3.90  \\
\bottomrule
\end{tabular}

\end{table*}

NIST assessors created the ground truth data that was used to evaluate Report Generation runs.
They did this using a process that largely follows the evaluation design described in \citet{mayfield2024evaluation}. 
A summary of the assessment statistics is presented in Table~\ref{tab:repgen-assessment-stats}. 
The first products of this assessment process were created together with the report requests.
After drafting a report request, the assessor manually wrote an example report.
Using their report request and their example report,
they then decided on the questions that a report responsive to the report request would need to answer.
At this time they also recorded any answers to each question that were known to them
from their research on the topic while writing their example report.
Answers were expressed in English rather than in the language of the document.
After an initial quality assurance check
these elements were finalized and retained for later use during evaluation.
Only the report request was provided to participating teams.
In total, we developed 59 report requests. We assign all requests to all three languages, resulting in 177 requests.

After runs for the Report Generation task were received,
assessors began ``nugget judgment.''
This process was timed to follow completion of relevance judgment for the CLIR and MLIR tasks
because it was useful to have the fullest possible set of relevant documents available.
The initial nugget questions developed immediately following the report request creation process
were then reviewed again and, when necessary, revised to ensure they were atomic and could be answered with phrases.
The questions were also marked as ``ok'' or ``vital'' during this review process.
A second quality assurance check was then performed.

Assessors then examined all relevant documents 
(based on relevance judgments for the associated CLIR topic)
and all cited documents (i.e., all documents cited in any submitted report).
The set of nugget answers was then revised to match the revised questions,
and extended based on additional relevant and cited documents that had been found.
Each answer was linked to all known documents that contained that answer. 

Concurrently with this answer revision process,
the assessors also judged all citations in each report sentence.
Citations were judged based on whether the facts expressed in the sentence containing the citation
were found in the cited document.
A sentence citation was scored as full, partial, or no support.
Reports could cite up to two documents per report sentence;
however, no attempt was made to determine if two partial scores
when combined would fully support the information in the sentence.
This was a limitation of the assessment process,
which collected all report sentences citing a particular document for the assessor to judge.
While sentences and citations were presented to assessors
without the context of the report from which they had been extracted,
assessors could access the entire report if they needed it to fully understand the information in a sentence.
At the end of this assessment phase,
a list of known correct answers to questions had been recorded
and all citations had been assessed.

The final assessment phase required the assessor to determine whether the sentence answered one or more nugget questions.
If it did answer a question,
the particular answer contained in the sentence was selected
or the answer was marked as ``other answer.''
If the sentence did answer one of the identified questions,
a sentence could be scored as
``other crucial nugget to the request''
indicating that an additional nugget question/answer pair should have been created (assuming the LLM did not hallucinate the information);
``topical nugget'' indicating the information was on topic but not necessary for the report;
``irrelevant nugget'' indicating a fact not responsive to the report request was included in the report;
and ``No nugget found'' indicating the sentence contained no facts
or its facts had previously been expressed in the report.
When performing this assessment phase,
assessors were not told whether a sentence under consideration contained citations.
This was meant to prevent biasing the question answering assessment
with effects from the presence or absence of suitable citations.
The number of supported sentences (supported by the citations) that answer a crucial or topical nugget that should have been created are summarized in Table~\ref{tab:repgen-assessment-stats}.

Both assessment processes were performed using the tool described in \citet{yang2025ragdemo}.
This tool facilitated linking answers to documents,
assessing citations, 
assessing whether report sentences contained facts,
and whether those facts were answers to a nugget question. 

Once citations were judged and answers to nugget questions had been identified,
those results were used to compute a score for each sentence.

\subsection{Additional Resources}

\subsubsection{Retrieval Service}

To support teams primarily focusing on the report generation task,
we provided a PLAID-X~\cite{tdistill} search service through a web API
that used an English-trained model~\cite{yang2024distillation}\footnote{\url{https://huggingface.co/hltcoe/plaidx-large-eng-tdist-mt5xxl-engeng}} for all languages. 
To minimize the resources needed to host the service,
we included the ability to remove documents in other than the requested language.. 
The user can request up to 100 documents for each query.
The service retrieves ten times the number of documents the user requested
to ensure enough documents in the requested language are retrieved. 

\subsubsection{Development Data}

Report generation topics were taken from the MLIR topic set.
NIST assessors were asked to generate report requests and sample reports for these topics.
Some topics were generated by more than one assessor,
both because assessors working on different languages shared topics,
and to support cross-assessor studies.
Track coordinators selected a single report request for each topic
for inclusion in the track data.
No software was available to support this process,
so assessors entered everything free-form in a shared document.
This led to a need for topic curation,
as citation and question formats differed across assessors.
Track coordinators manually curated the report requests,
and converted them to JSON format.
A total of forty-seven report requests were released as development data,
statistics on which appear in Table~\ref{tab:devr}.
Because the questions and answers were not normalized,
questions were issued with the following caution:
``Warning: these questions and answers are not all representative of the way 
questions will be asked and answered in the pilot,
because some questions are compound and some answers do not include citations.''

\begin{table*}[]
\caption{Development report request statistics. }\label{tab:devr}
    \centering
    
\begin{tabular}{l|ccc}
\toprule
 & Chinese & Persian & Russian  \\
\midrule
\#  Reports       &   19    &  8     &  20      \\
Avg. \# Characters / Reports &   1865.2    &   2256.0    &   1894.1     \\
Avg. \# Sentences / Reports &   15.0    &   16.8    &   16.6   \\
Avg. \# Citations / Reports &   10.0    &   11.1    &   16.9   \\
Avg. \# Unique Citations / Reports &   6.3    &   5.9    &   7.8   \\
Avg. \# Questions / Reports &   11.5    &   12.4    &   14.4    \\
\bottomrule
\end{tabular}

\end{table*}

\subsection{Participation}

We received 51 runs for the Report Generation task from four participating teams,
including 17 runs for each of the three document languages:
\begin{itemize}
    \item Johns Hopkins University HLTCOE (4 runs per language)~\cite{neuclir-participants-hltcoe}
    \item University of Waterloo (4 runs per language)~\cite{neuclir-participants-h2oloo}
    \item University of Amsterdam (7 runs per language)~\cite{neuclir-participants-ams}
    \item IDA/CCS (2 runs per language)~\cite{neuclir-participants-ida}
\end{itemize}

The track coordinators did not prepare baseline runs for this pilot task.  %
\subsection{Results}

\begin{figure*}
    \centering
    \includegraphics[width=\linewidth]{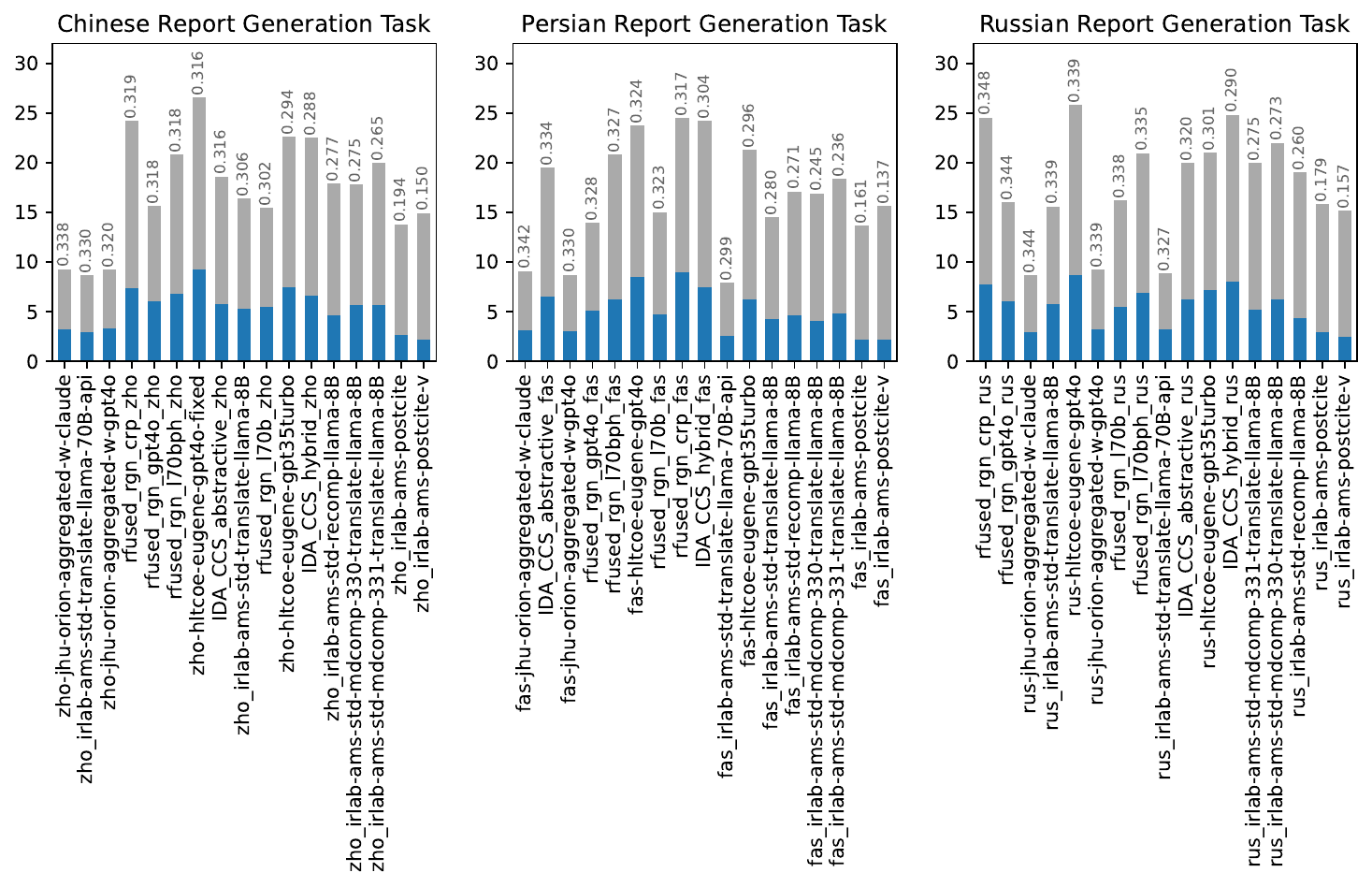}
    \caption{Bar chart of the number of relevant citations (in blue)
    across all cited documents (in gray) for each submitted run. 
    Each bar represents a run. 
    Values at the top of each bar are the document-level precision of the generated report,
    which are reported in Tables~\ref{tab:zho-repgen-results}, \ref{tab:fas-repgen-results}, and \ref{tab:rus-repgen-results}. 
    }\label{fig:repgen_citation_rel_bar}
\end{figure*}

\begin{figure*}
    \centering
    \includegraphics[width=\linewidth]{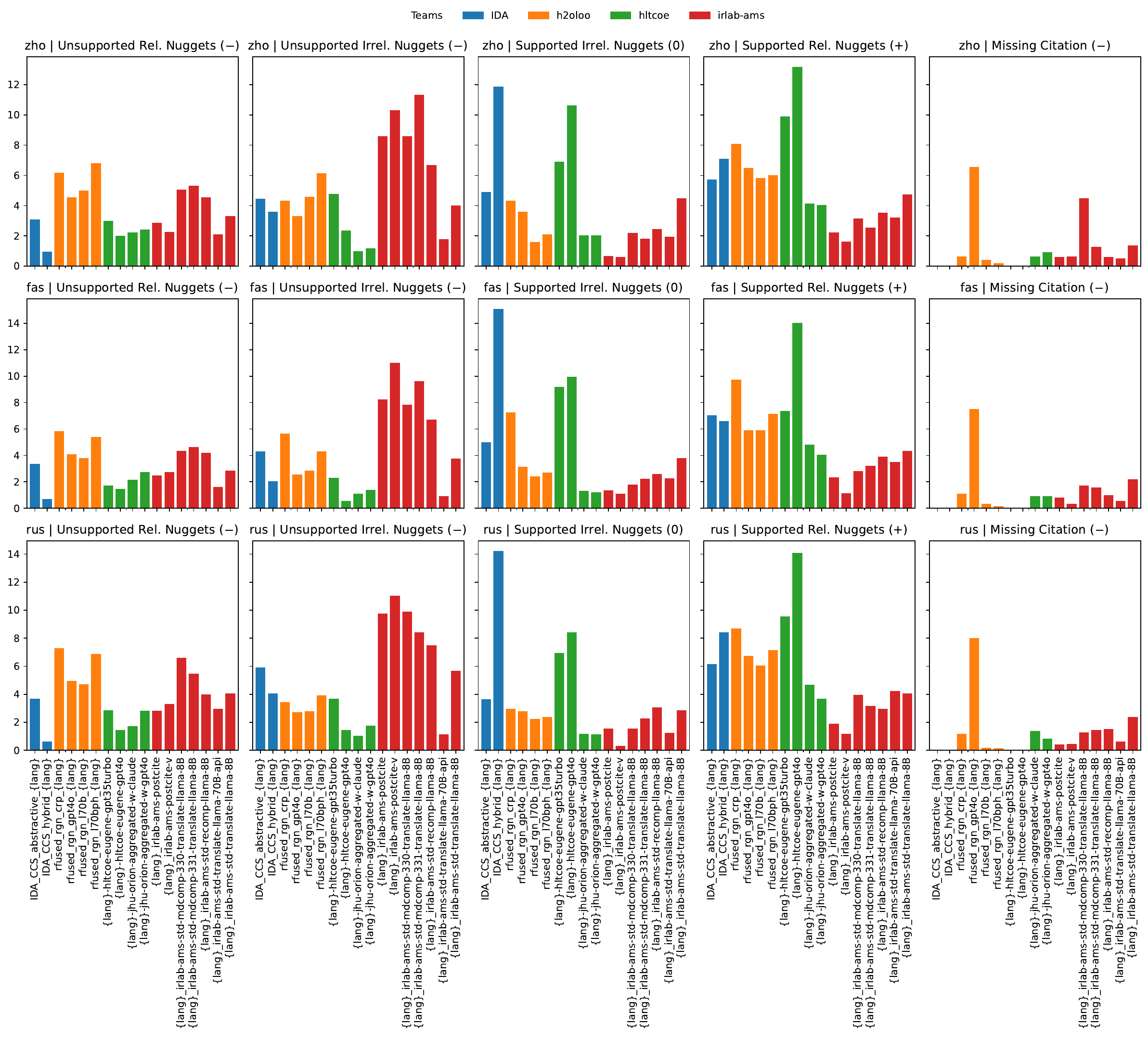}
    \caption{Components of ARGUE scores for each Report Generation run. }\label{fig:repgen_argue_breakdown}
\end{figure*}

\begin{table}[]

\caption{Relevance agreement between ranked retrieval (qrels) and nugget annotation.}\label{tab:qrels_vs_nugget}
\centering
\begin{tabular}{l|rr|rr|rr}
\toprule
               & \multicolumn{6}{c}{From Nugget Annotation} \\
\midrule
               & \multicolumn{2}{c|}{Chinese} & \multicolumn{2}{c|}{Persian} & \multicolumn{2}{r}{Russian} \\
From qrels     & NRel. & Rel. & NRel. & Rel. & NRel. & Rel.  \\
\midrule
Non-Relevant   & 475 & 372 & 480 & 310 & 455 & 424 \\
    Relevant   &  37 & 169 &  24 & 249 &  35 & 278 \\
\bottomrule
\end{tabular}

\end{table}

Tables~\ref{tab:zho-repgen-results}, \ref{tab:fas-repgen-results}, and \ref{tab:rus-repgen-results}
show ARGUE report generation scores for Chinese, Persian, and Russian respectively.
Figure~\ref{fig:repgen_argue_breakdown} indicates how the scores in those tables were derived.
Scores were mapped to a zero-to-one scale by scoring categories labeled ($-$) as 0.0
and the single category labeled (+), Supported Relevant Nuggets, as 1.0. 
Neutral labels (0) were ignored
by removing them from both numerators and denominators of the ARGUE precision calculation. 

\subsubsection{Citation Precision}
is the proportion of citations that accurately refer to a relevant document.
There are two ways a document might be considered relevant for this purpose:
the relevance label is derived from nugget annotation
(that is, any document that contains a mention of a nugget is treated as relevant),
or by using the qrels for ranked retrieval
(shown in Table~\ref{tab:qrels_vs_nugget}).
Despite the apparent difference in the definition of relevance
(or approach for acquiring such annotations),
Pearson's rank correlations under the two definitions are high -- 0.9 for all three languages.
    
\subsubsection{Nugget Recall and Support}

\textit{Nugget recall} is the proportion of nugget questions correctly answered by at least one report sentence.
Multiple report sentences correctly answering a single nugget question are counted only once.
In contrast, \textit{nugget support}
is the proportion of the report's nuggets for which the sentence that answers a nugget question
is supported by its cited documents.
Both measures require the nugget mentioned in the report to be supported by the cited documents to be credited.

\subsubsection{Sentence Support}
Similar to Nugget Support, \textit{sentence support} is the proportion of sentences in a report
supported by its cited documents.
This includes support for information that is not captured by nuggets
(which is unlikely to be relevant to the report request).
Such information is allowed in a report without penalty
to partially address disagreements between systems and assessors
over which nuggets are crucial to a report.

\subsubsection{Scores}

The top submitted runs score close to 0.9 on the main ARGUE measure.
However, top citation precision scores are only around 0.3,
suggesting that systems can still improve on the inclusion of relevant information in the report.
Top nugget recall is less than 0.5,
indicating that there is also room for improvement in nugget/fact coverage of the topic.

\section{Future Directions}

The NeuCLIR track is retiring after TREC 2024.
In 2025, we will transition to the new RAGTIME track,
where the primary task will be report generation from multilingual news content
in Arabic, Chinese, English, and Russian.

\subsection{What's New in RAGTIME 2025?}

RAGTIME will include a new test collection,
with the size of each of the four language components balanced across those languages. 
Report Generation in RAGTIME will differ from our early work with Report generation in NeuCLIR in three important ways.
First, the report request in RAGTIME is generated \emph{de novo} for the task,
rather than having been developed based on a topic originally developed for MLIR (as was the case in NeuCLIR).
Second, the RAGTIME report generation task is not limited to documents in a single language.
Instead, systems will be asked to draw on content in all four RAGTIME languages.
Notably, one of the languages is English, the same language as the report request and the report.
Thus RAGTIME involves the additional challenge of integrating same-language and cross-language sources,
which was not present even in the NeuCLIR MLIR task
(because the NeuCLIR test collection did not include English).
Third, RAGTIME will also include an emphasis on automating parts of the evaluation,
with an eye toward fostering reusability of the test collection.

Our present plans for RAGTIME also include four changes %
from NeuCLIR that are intended to facilitate the entry of new participants.
First, the RAGTIME collections are somewhat smaller than those of NeuCLIR, with about 1 million documents per language. 
This change is intended to deemphasize the scalability of the retrieval component
to allow participating teams to focus more on report generation. 
Second, because two of the source languages will be new,
the TREC 2025 RAGTIME track will include an early-summer dry run
to give participating teams access to some manually annotated development data for their final systems.
Third, there will be a monolingual English task for teams that do not want to work with non-English content.
Fourth, the RAGTIME output formats are consistent with those of other TREC RAG tracks
(BioGen, DRAGUN, and RAG)
to simplify participation in several of those tracks.

We are not currently planning future work with the NeuCLIR Technical Documents collection
because that collection lacks the degree of topical convergence across documents that we believe would be needed
for use in a Report Generation task.
We believe that the collection is suitable in its present form
for evaluating the ability of a CLIR system to handle technical vocabulary.

\subsection{What's Not Changing?}

There will be some continuity along with these changes.
Most notably, RAGTIME will continue to report results for CLIR and MLIR for teams who have interest in those tasks.
This benefits reusability of the RAGTIME data through pool enrichment,
because CLIR and MLIR runs may find documents that no Report Generation system chose to include in its report.
We expect this will be of particular interest for Arabic,
where we expect RAGTIME to produce the largest available CLIR test collection in that language. 
Notably, however, the form and content of the topics will differ in RAGTIME.
Specifically, while there will be a title field
(which can be used as a short web-style query), %
the traditional TREC description and narrative fields will be replaced with the report request,
and relevance judgments will be based on that report request.
Thus the RAGTIME CLIR and MLIR tasks may help to push the state of the art for retrieval over very long queries.

\section{Conclusion}

In this third and final iteration of the TREC NeuCLIR track,
we have completed our development of the first set of large CLIR test collections
with judgment pools augmented by modern transformer-based neural information retrieval systems.
This NeuCLIR news collection includes more than 100 topics for each of the three languages.
As in prior years, we continue to see strongest effectiveness from neural CLIR systems.  
The NeuCLIR test collection also supports MLIR evaluation,
which continues to be a more challenging task for systems than is CLIR.

NeuCLIR 2024 was the second and final year of the Chinese Technical Documents CLIR task,
which has produced a new test collection that now also has more than 100 topics.
In contrast to 2023 when no training data specific to this task had been available,
we saw the emergence of relatively strong CLIR systems for these more challenging documents
that are not reliant on machine translation of the full document set.

Finally, NeuCLIR 2024 has also served as an incubator for a new Report Generation task,
which has a bright future.
Results on the NeuCLIR Report Generation task are promising,
but reveal areas that need significant further research.
Experience with the design of the NeuCLIR task
has informed the design of the TREC 2025 RAGTIME track.  

In short, the three years of the TREC NeuCLIR achieved what it set out to do, and more.
While the test collections built by the track are one clear legacy with enduring value,
it is the research community that has and will make use of those test collections
that we expect will be the track's most lasting legacy.

\bibliographystyle{ACM-Reference-Format}
\bibliography{biblio}

\appendix

\begin{figure*}
    \centering
    \includegraphics[width=0.85\linewidth]{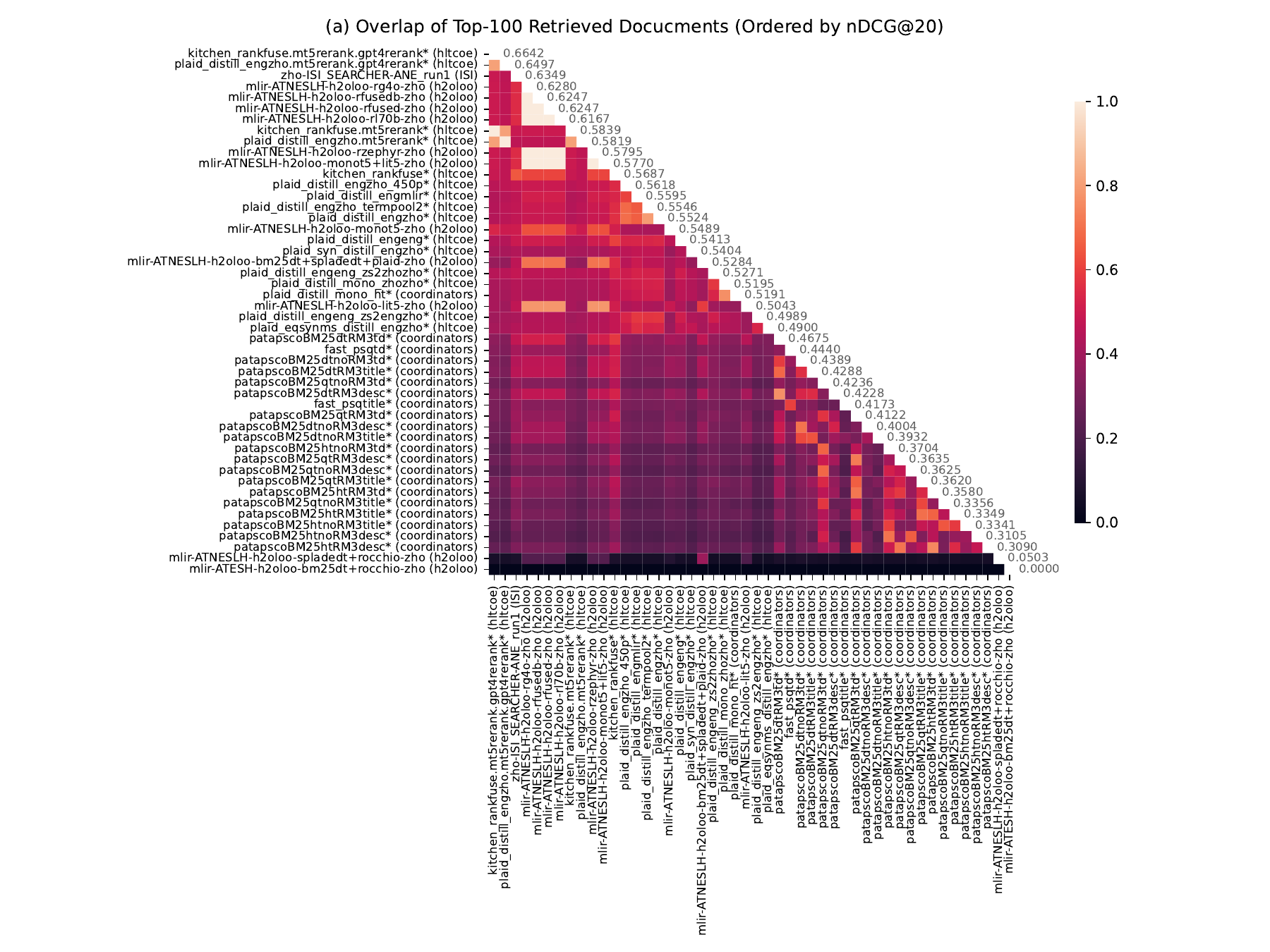}
    \includegraphics[width=0.85\linewidth]{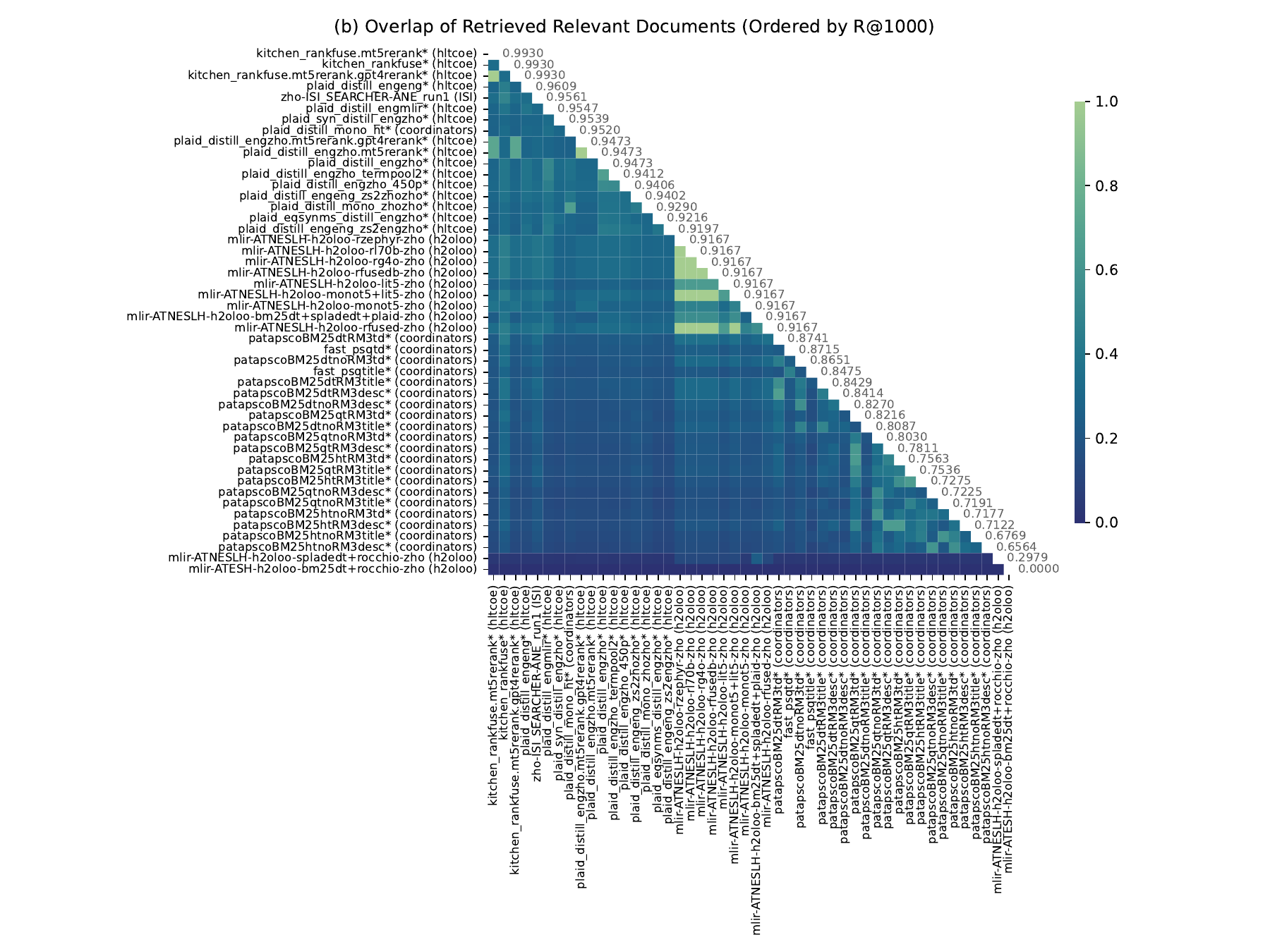}
    \caption{Overlap of documents retrieved by systems that participated in Chinese. * indicates manual runs.}
    \label{fig:zho-overlap}
\end{figure*}

\begin{figure*}
    \centering
    \includegraphics[width=0.85\linewidth]{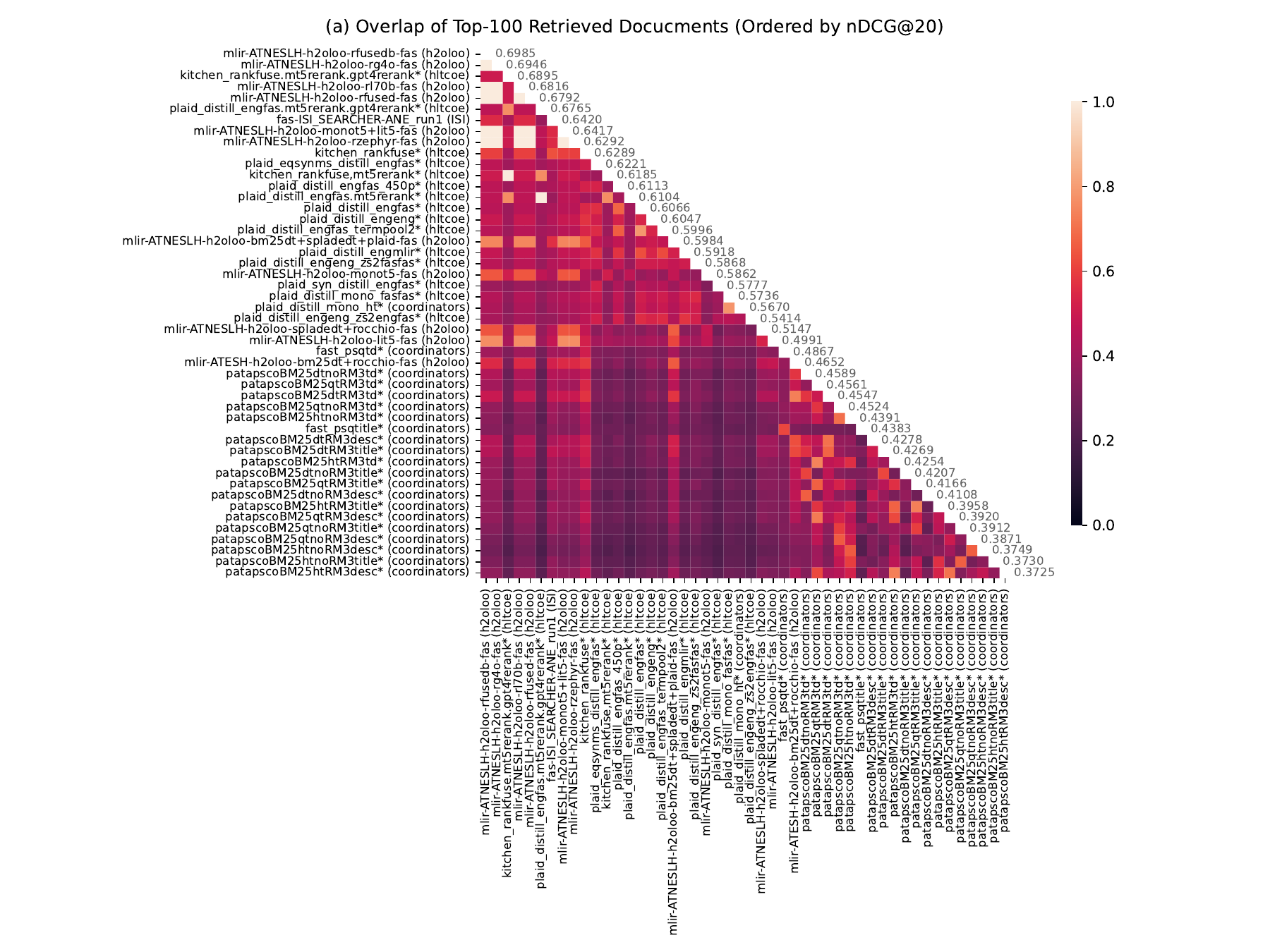}
    \includegraphics[width=0.85\linewidth]{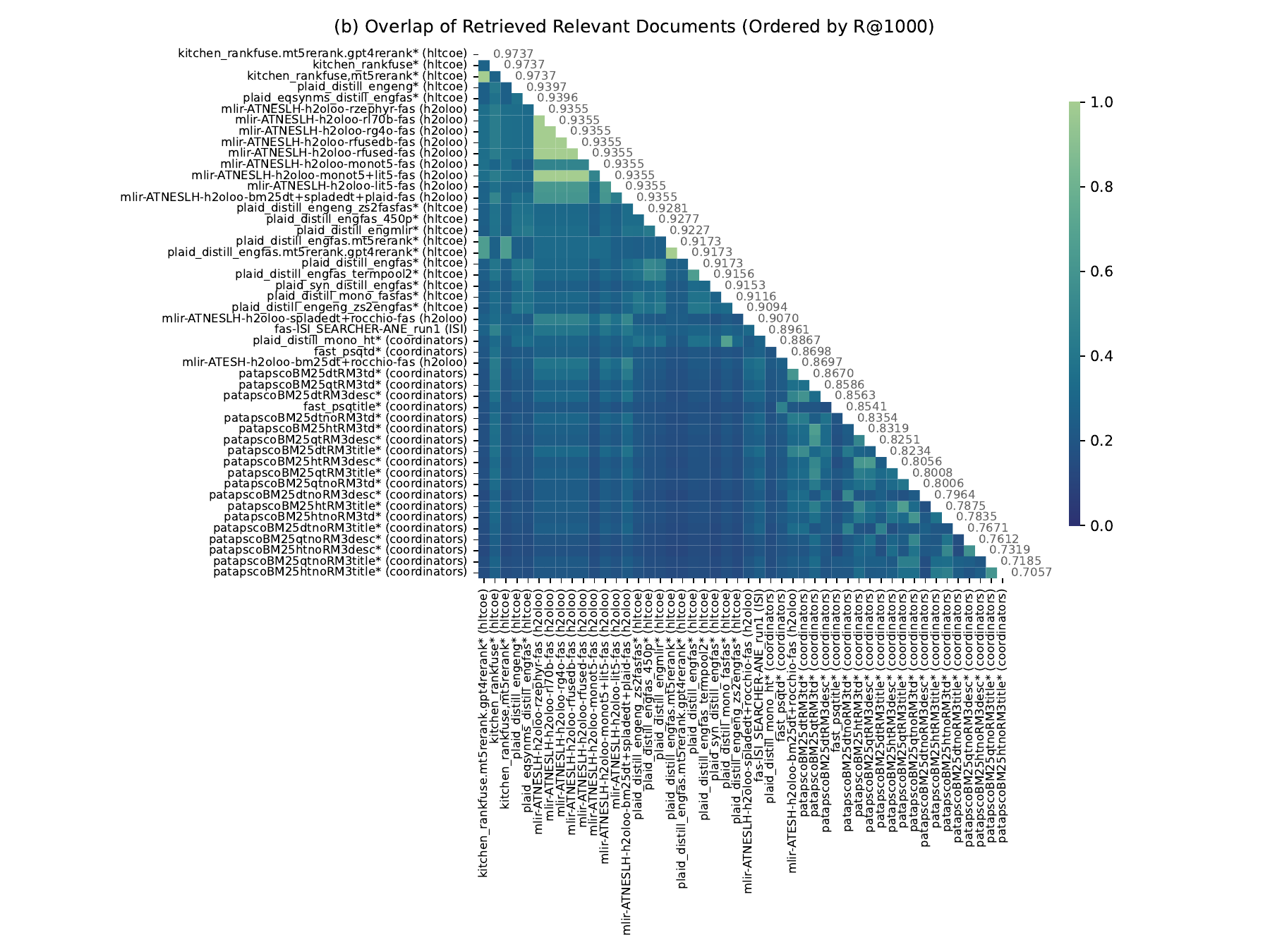}
    \caption{Overlap of documents retrieved by systems that participated in Persian. * indicates manual runs.}
    \label{fig:fas-overlap}
\end{figure*}

\begin{figure*}
    \centering
    \includegraphics[width=0.85\linewidth]{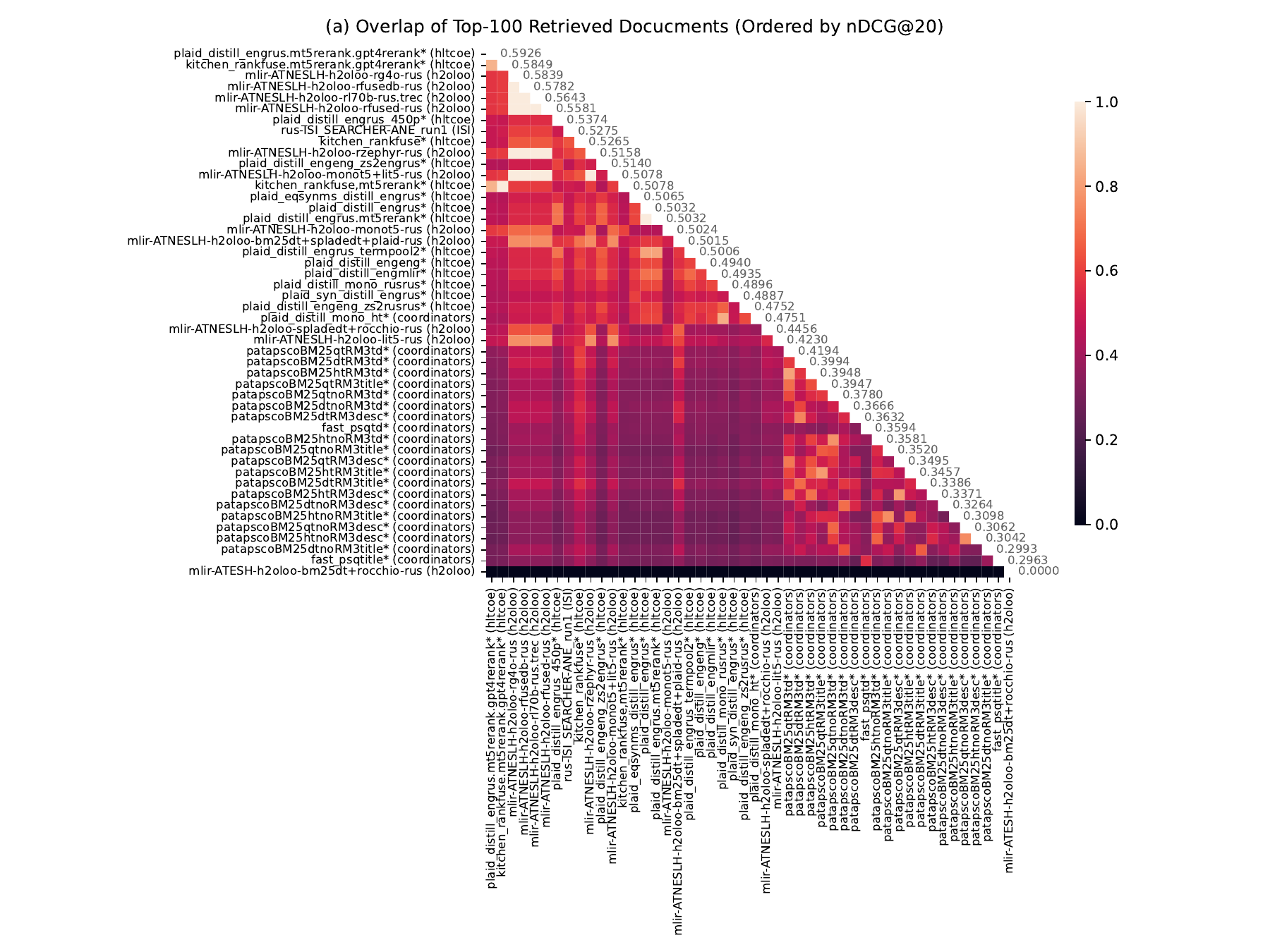}
    \includegraphics[width=0.85\linewidth]{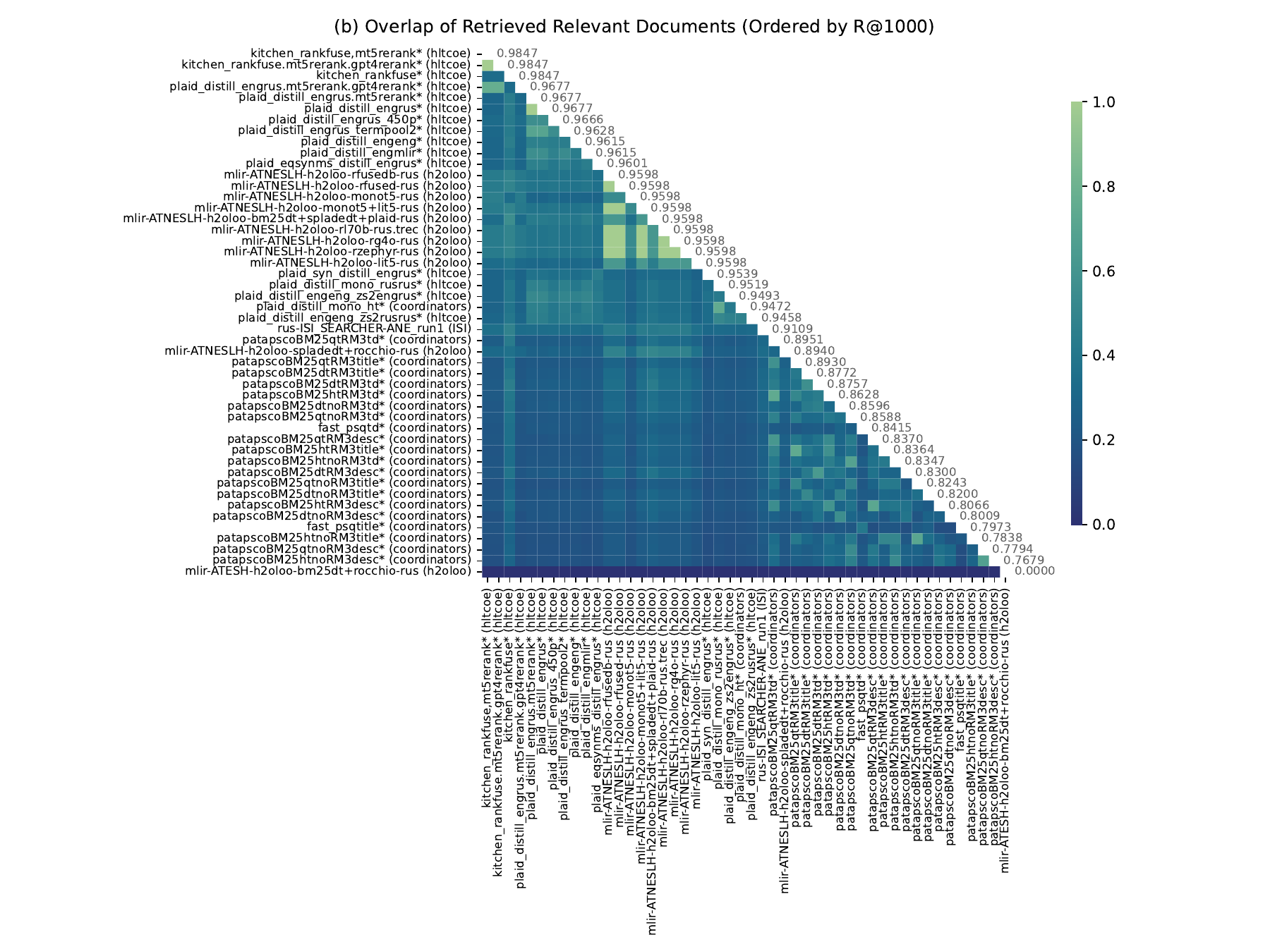}
    \caption{Overlap of documents retrieved by systems that participated in Russian. * indicates manual runs.}
    \label{fig:rus-overlap}
\end{figure*}

\begin{figure*}
    \centering
    \includegraphics[width=0.85\linewidth]{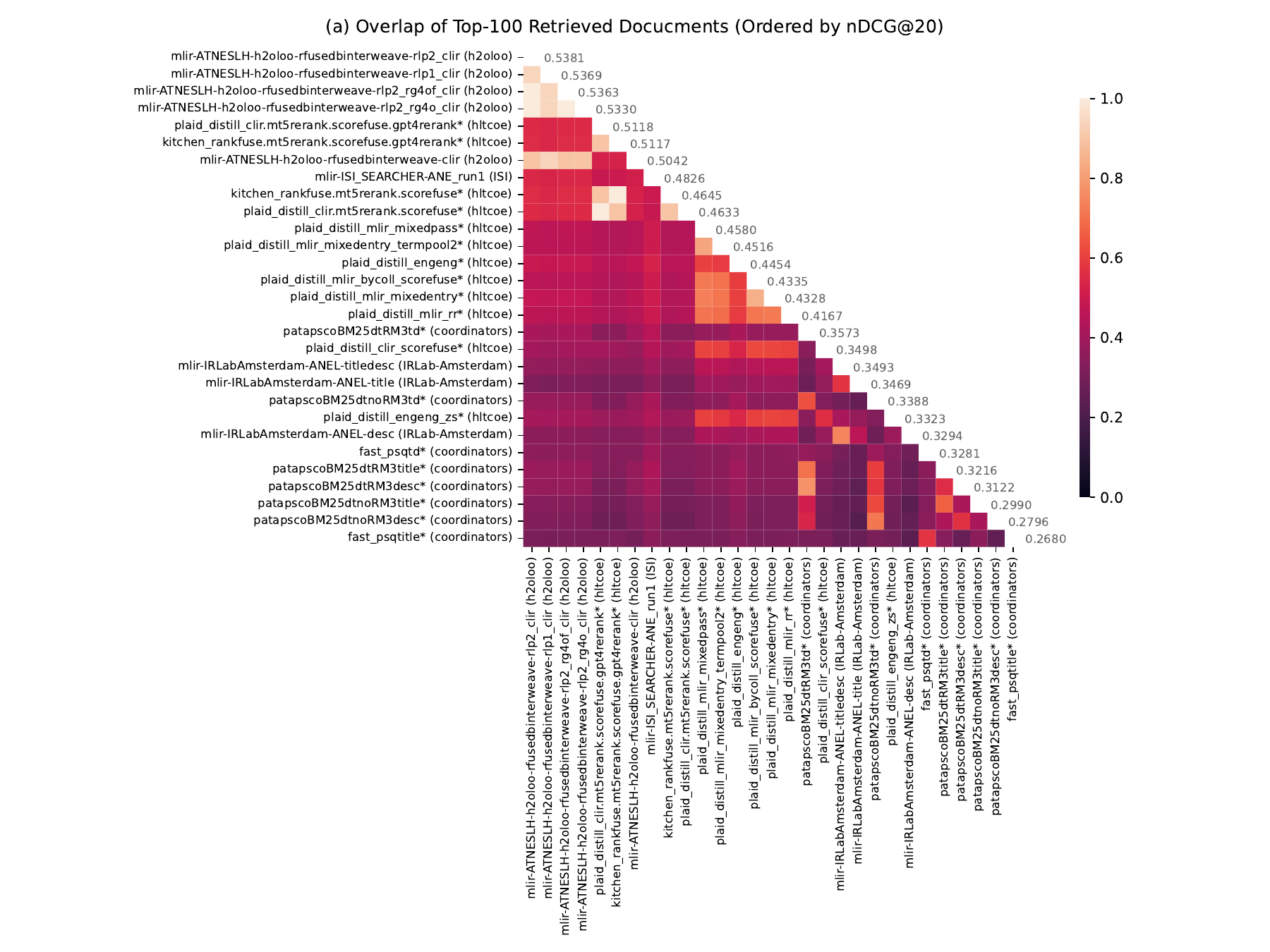}
    \includegraphics[width=0.85\linewidth]{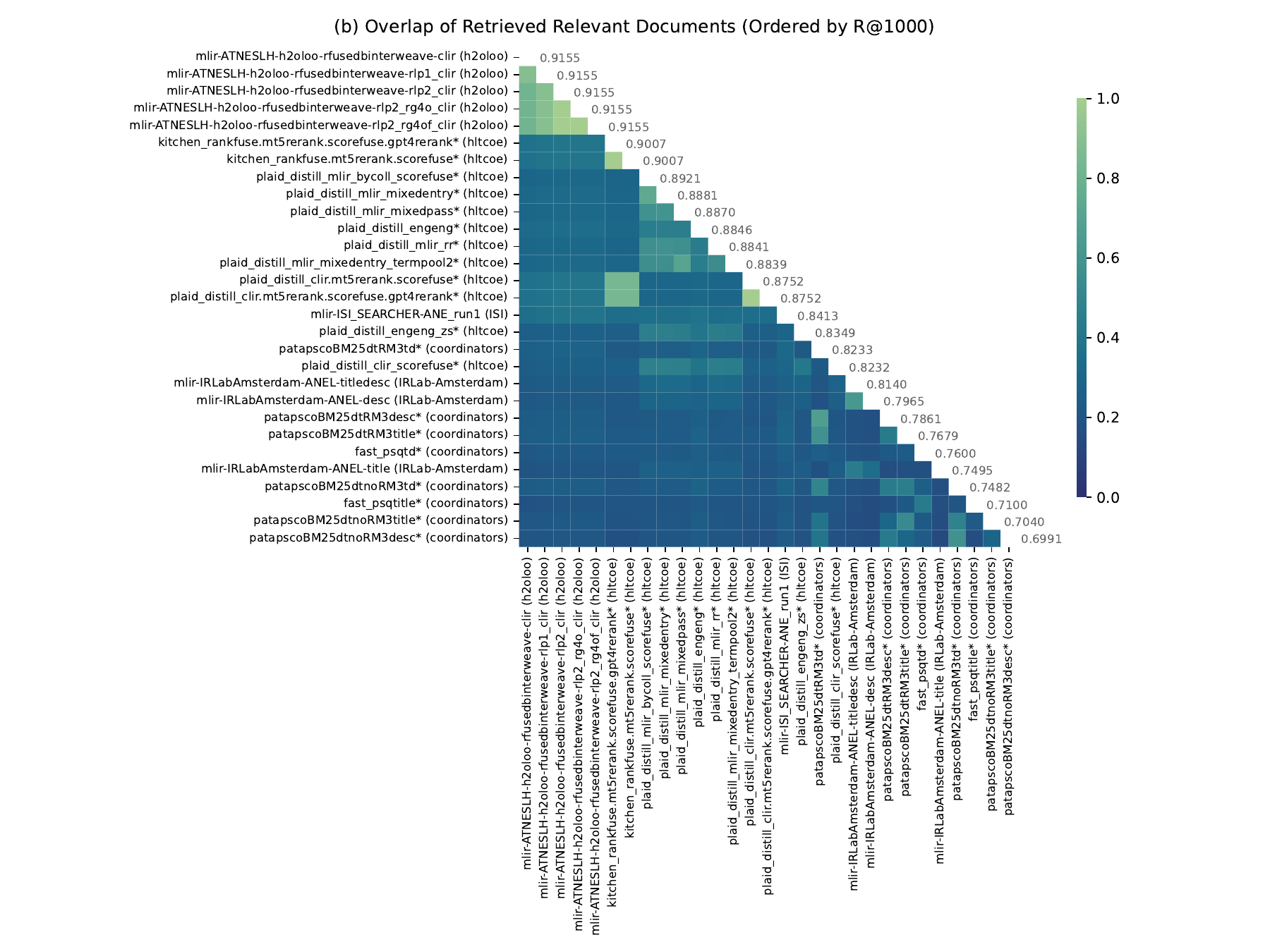}
    \caption{Overlap of documents retrieved by systems that participated in MLIR runs. * indicates manual runs.}
    \label{fig:mlir-overlap}
\end{figure*}

\begin{figure*}
    \centering
    \includegraphics[width=0.85\linewidth]{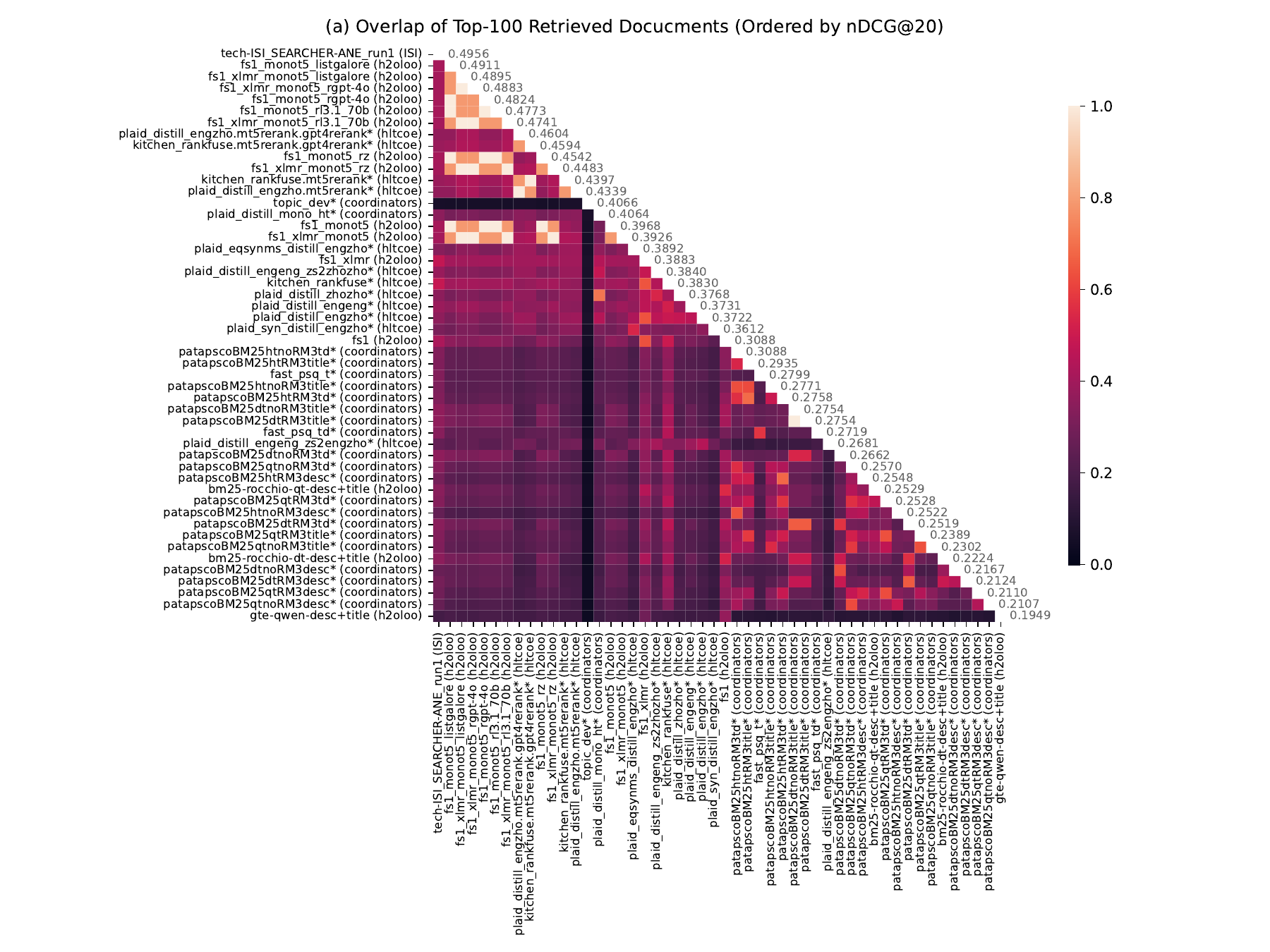}
    \includegraphics[width=0.85\linewidth]{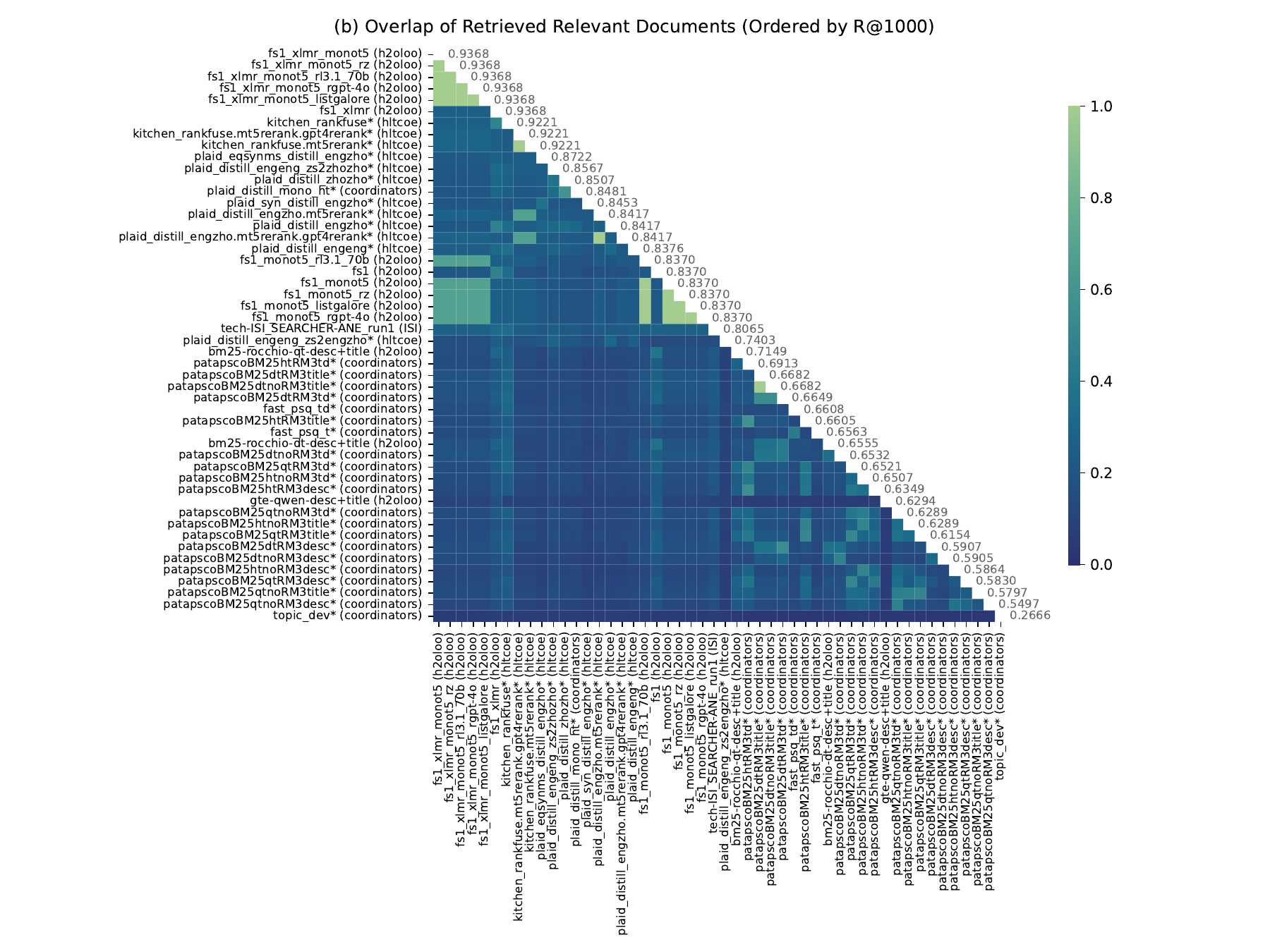}
    \caption{Overlap of documents retrieved by systems that participated in the Technical Document Task. * indicates manual runs.}
    \label{fig:tech-overlap}
\end{figure*}

\begin{table*}
\setlength\tabcolsep{0.4em}

\caption{Chinese Results.
Monolingual runs, which use human translations of the queries, are shown in green.
* indicates manual runs. 
Column ``JFD'' indicates whether the run is \underline{j}udged at \underline{f}ull \underline{d}epth, which is 100.
Other runs were judged to depth 50. 
}\label{tab:zho-full-results}
\centering
\begin{tabular}{ll|ccc|ccccc}
\toprule
        Team &                                      Run Name &    JFD &       DS &         QS &  nDCG@20 &   RBP &    AP &  R@100 &  R@1k \\
\midrule
      hltcoe &       kitchen\_rankfuse.mt5rerank.gpt4rerank* & \cmark &  Orig+DT &    Orig+GT &    0.664 & 0.524 & 0.578 &  0.836 & 0.993 \\
      hltcoe &  plaid\_distill\_engzho.mt5rerank.gpt4rerank* & \cmark &     Orig &       Orig &    0.650 & 0.503 & 0.541 &  0.808 & 0.947 \\
         ISI &                   zho-ISI\_SEARCHER-ANE\_run1 & \cmark &     Orig &       Orig &    0.635 & 0.480 & 0.540 &  0.881 & 0.956 \\
      h2oloo &                  mlir-ATNESLH-h2oloo-rg4o-zho & \cmark &  Orig+DT &    Orig+GT &    0.628 & 0.487 & 0.537 &  0.814 & 0.917 \\
      h2oloo &               mlir-ATNESLH-h2oloo-rfusedb-zho & \cmark &  Orig+DT &    Orig+GT &    0.625 & 0.488 & 0.533 &  0.814 & 0.917 \\
      h2oloo &                mlir-ATNESLH-h2oloo-rfused-zho & \cmark &  Orig+DT &    Orig+GT &    0.625 & 0.483 & 0.530 &  0.814 & 0.917 \\
      h2oloo &                 mlir-ATNESLH-h2oloo-rl70b-zho & \cmark &  Orig+DT &    Orig+GT &    0.617 & 0.482 & 0.526 &  0.814 & 0.917 \\
      hltcoe &                  kitchen\_rankfuse.mt5rerank* & \cmark &  Orig+DT &    Orig+GT &    0.584 & 0.450 & 0.502 &  0.836 & 0.993 \\
      hltcoe &             plaid\_distill\_engzho.mt5rerank* & \cmark &     Orig &       Orig &    0.582 & 0.445 & 0.485 &  0.808 & 0.947 \\
      h2oloo &               mlir-ATNESLH-h2oloo-rzephyr-zho & \cmark &  Orig+DT &    Orig+GT &    0.580 & 0.455 & 0.486 &  0.814 & 0.917 \\
      h2oloo &           mlir-ATNESLH-h2oloo-monot5+lit5-zho & \cmark &  Orig+DT &    Orig+GT &    0.577 & 0.455 & 0.488 &  0.814 & 0.917 \\
      hltcoe &                            kitchen\_rankfuse* & \cmark &  Orig+DT &    Orig+GT &    0.569 & 0.436 & 0.497 &  0.852 & 0.993 \\
      hltcoe &                 plaid\_distill\_engzho\_450p* & \xmark &     Orig &       Orig &    0.562 & 0.438 & 0.488 &  0.795 & 0.941 \\
      hltcoe &                      plaid\_distill\_engmlir* & \xmark &     Orig &       Orig &    0.560 & 0.422 & 0.472 &  0.812 & 0.955 \\
      hltcoe &            plaid\_distill\_engzho\_termpool2* & \xmark &     Orig &       Orig &    0.555 & 0.416 & 0.471 &  0.825 & 0.941 \\
      hltcoe &                       plaid\_distill\_engzho* & \cmark &     Orig &       Orig &    0.552 & 0.421 & 0.472 &  0.811 & 0.947 \\
      h2oloo &                mlir-ATNESLH-h2oloo-monot5-zho & \cmark &  Orig+DT &    Orig+GT &    0.549 & 0.440 & 0.460 &  0.793 & 0.917 \\
      hltcoe &                       plaid\_distill\_engeng* & \xmark &       DT &       Orig &    0.541 & 0.420 & 0.467 &  0.823 & 0.961 \\
      hltcoe &                  plaid\_syn\_distill\_engzho* & \cmark &     Orig &       Orig &    0.540 & 0.387 & 0.454 &  0.815 & 0.954 \\
      h2oloo & mlir-ATNESLH-h2oloo-bm25dt+spladedt+plaid-zho & \cmark &  Orig+DT &    Orig+GT &    0.528 & 0.389 & 0.439 &  0.754 & 0.917 \\
      hltcoe &            plaid\_distill\_engeng\_zs2zhozho* & \xmark &     Orig &         GT &    0.527 & 0.392 & 0.445 &  0.824 & 0.940 \\
      hltcoe &                 plaid\_distill\_mono\_zhozho* & \xmark &     Orig &         GT &    0.520 & 0.382 & 0.433 &  0.789 & 0.929 \\
coordinators &                \ml{plaid\_distill\_mono\_ht*} & \cmark &     Orig &         HT &    0.519 & 0.402 & 0.452 &  0.831 & 0.952 \\
      h2oloo &                  mlir-ATNESLH-h2oloo-lit5-zho & \cmark &  Orig+DT &    Orig+GT &    0.504 & 0.376 & 0.410 &  0.776 & 0.917 \\
      hltcoe &            plaid\_distill\_engeng\_zs2engzho* & \cmark &     Orig &       Orig &    0.499 & 0.380 & 0.407 &  0.761 & 0.920 \\
      hltcoe &              plaid\_eqsynms\_distill\_engzho* & \cmark &     Orig &       Orig &    0.490 & 0.377 & 0.426 &  0.759 & 0.922 \\
coordinators &                          patapscoBM25dtRM3td* & \cmark &       DT &       Orig &    0.468 & 0.368 & 0.395 &  0.688 & 0.874 \\
coordinators &                                  fast\_psqtd* & \cmark &     Orig &       Orig &    0.444 & 0.337 & 0.363 &  0.701 & 0.871 \\
coordinators &                        patapscoBM25dtnoRM3td* & \cmark &       DT &       Orig &    0.439 & 0.338 & 0.358 &  0.667 & 0.865 \\
coordinators &                       patapscoBM25dtRM3title* & \xmark &       DT &       Orig &    0.429 & 0.324 & 0.361 &  0.671 & 0.843 \\
coordinators &                        patapscoBM25qtnoRM3td* & \cmark &     Orig &         GT &    0.424 & 0.321 & 0.316 &  0.602 & 0.803 \\
coordinators &                        patapscoBM25dtRM3desc* & \xmark &       DT &       Orig &    0.423 & 0.327 & 0.348 &  0.648 & 0.841 \\
coordinators &                               fast\_psqtitle* & \xmark &     Orig &       Orig &    0.417 & 0.314 & 0.331 &  0.677 & 0.848 \\
coordinators &                          patapscoBM25qtRM3td* & \cmark &     Orig &         GT &    0.412 & 0.333 & 0.348 &  0.661 & 0.822 \\
coordinators &                      patapscoBM25dtnoRM3desc* & \xmark &       DT &       Orig &    0.400 & 0.301 & 0.312 &  0.609 & 0.827 \\
coordinators &                     patapscoBM25dtnoRM3title* & \xmark &       DT &       Orig &    0.393 & 0.300 & 0.307 &  0.604 & 0.809 \\
coordinators &                   \ml{patapscoBM25htnoRM3td*} & \cmark &     Orig &         HT &    0.370 & 0.290 & 0.289 &  0.524 & 0.718 \\
coordinators &                        patapscoBM25qtRM3desc* & \xmark &     Orig &         GT &    0.363 & 0.272 & 0.304 &  0.594 & 0.781 \\
coordinators &                      patapscoBM25qtnoRM3desc* & \xmark &     Orig &         GT &    0.363 & 0.257 & 0.274 &  0.545 & 0.723 \\
coordinators &                       patapscoBM25qtRM3title* & \xmark &     Orig &         GT &    0.362 & 0.275 & 0.298 &  0.571 & 0.754 \\
coordinators &                     \ml{patapscoBM25htRM3td*} & \cmark &     Orig &         HT &    0.358 & 0.302 & 0.306 &  0.560 & 0.756 \\
coordinators &                     patapscoBM25qtnoRM3title* & \xmark &     Orig &         GT &    0.336 & 0.250 & 0.250 &  0.489 & 0.719 \\
coordinators &                  \ml{patapscoBM25htRM3title*} & \cmark &     Orig &         HT &    0.335 & 0.279 & 0.287 &  0.541 & 0.728 \\
coordinators &                \ml{patapscoBM25htnoRM3title*} & \xmark &     Orig &         HT &    0.334 & 0.265 & 0.258 &  0.487 & 0.677 \\
coordinators &                 \ml{patapscoBM25htnoRM3desc*} & \xmark &     Orig &         HT &    0.311 & 0.226 & 0.246 &  0.483 & 0.656 \\
coordinators &                   \ml{patapscoBM25htRM3desc*} & \xmark &     Orig &         HT &    0.309 & 0.242 & 0.261 &  0.517 & 0.712 \\
      h2oloo &      mlir-ATNESLH-h2oloo-spladedt+rocchio-zho & \xmark &  Orig+DT &    Orig+GT &    0.050 & 0.037 & 0.036 &  0.112 & 0.298 \\
      h2oloo &          mlir-ATESH-h2oloo-bm25dt+rocchio-zho & \xmark &  Orig+DT &    Orig+GT &    0.000 & 0.000 & 0.000 &  0.000 & 0.000 \\
\bottomrule
\end{tabular}

\end{table*}

\begin{table*}
\setlength\tabcolsep{0.5em}

\caption{Persian Results.
Monolingual runs, which use human translations of the queries, are marked as green.
* indicates manual runs.
Column ``JFD'' indicates whether the run is \underline{j}udged at \underline{f}ull \underline{d}epth, which is 100.
Other runs were judged to depth 50. 
}\label{tab:fas-full-results}
    \centering

\centering
\begin{tabular}{ll|ccc|ccccc}
\toprule
        Team &                                      Run Name &    JFD &     DS &         QS &  nDCG@20 &   RBP &    AP &  R@100 &  R@1k \\
\midrule
      h2oloo &               mlir-ATNESLH-h2oloo-rfusedb-fas & \cmark &  Orig+DT &    Orig+GT &    0.698 & 0.572 & 0.611 &  0.790 & 0.936 \\
      h2oloo &                  mlir-ATNESLH-h2oloo-rg4o-fas & \cmark &  Orig+DT &    Orig+GT &    0.695 & 0.576 & 0.616 &  0.790 & 0.936 \\
      hltcoe &       kitchen\_rankfuse.mt5rerank.gpt4rerank* & \cmark &  Orig+DT &    Orig+GT &    0.690 & 0.570 & 0.583 &  0.782 & 0.974 \\
      h2oloo &                 mlir-ATNESLH-h2oloo-rl70b-fas & \cmark &  Orig+DT &    Orig+GT &    0.682 & 0.562 & 0.594 &  0.790 & 0.936 \\
      h2oloo &                mlir-ATNESLH-h2oloo-rfused-fas & \cmark &  Orig+DT &    Orig+GT &    0.679 & 0.569 & 0.599 &  0.790 & 0.936 \\
      hltcoe &  plaid\_distill\_engfas.mt5rerank.gpt4rerank* & \cmark &     Orig &       Orig &    0.676 & 0.569 & 0.572 &  0.766 & 0.917 \\
         ISI &                   fas-ISI\_SEARCHER-ANE\_run1 & \cmark &     Orig &       Orig &    0.642 & 0.517 & 0.550 &  0.781 & 0.896 \\
      h2oloo &           mlir-ATNESLH-h2oloo-monot5+lit5-fas & \cmark &  Orig+DT &    Orig+GT &    0.642 & 0.518 & 0.560 &  0.790 & 0.936 \\
      h2oloo &               mlir-ATNESLH-h2oloo-rzephyr-fas & \cmark &  Orig+DT &    Orig+GT &    0.629 & 0.544 & 0.557 &  0.790 & 0.936 \\
      hltcoe &                            kitchen\_rankfuse* & \cmark &  Orig+DT &    Orig+GT &    0.629 & 0.485 & 0.547 &  0.789 & 0.974 \\
      hltcoe &              plaid\_eqsynms\_distill\_engfas* & \cmark &     Orig &       Orig &    0.622 & 0.472 & 0.552 &  0.791 & 0.940 \\
      hltcoe &                  kitchen\_rankfuse,mt5rerank* & \cmark &  Orig+DT &    Orig+GT &    0.619 & 0.500 & 0.526 &  0.782 & 0.974 \\
      hltcoe &                 plaid\_distill\_engfas\_450p* & \xmark &     Orig &       Orig &    0.611 & 0.480 & 0.543 &  0.774 & 0.928 \\
      hltcoe &             plaid\_distill\_engfas.mt5rerank* & \cmark &     Orig &       Orig &    0.610 & 0.494 & 0.516 &  0.766 & 0.917 \\
      hltcoe &                       plaid\_distill\_engfas* & \cmark &     Orig &       Orig &    0.607 & 0.469 & 0.536 &  0.775 & 0.917 \\
      hltcoe &                       plaid\_distill\_engeng* & \xmark &       DT &       Orig &    0.605 & 0.474 & 0.523 &  0.780 & 0.940 \\
      hltcoe &            plaid\_distill\_engfas\_termpool2* & \xmark &     Orig &       Orig &    0.600 & 0.466 & 0.528 &  0.763 & 0.916 \\
      h2oloo & mlir-ATNESLH-h2oloo-bm25dt+spladedt+plaid-fas & \cmark &  Orig+DT &    Orig+GT &    0.598 & 0.468 & 0.513 &  0.776 & 0.936 \\
      hltcoe &                      plaid\_distill\_engmlir* & \xmark &     Orig &       Orig &    0.592 & 0.451 & 0.511 &  0.759 & 0.923 \\
      hltcoe &            plaid\_distill\_engeng\_zs2fasfas* & \xmark &     Orig &         GT &    0.587 & 0.455 & 0.529 &  0.770 & 0.928 \\
      h2oloo &                mlir-ATNESLH-h2oloo-monot5-fas & \cmark &  Orig+DT &    Orig+GT &    0.586 & 0.490 & 0.508 &  0.778 & 0.936 \\
      hltcoe &                  plaid\_syn\_distill\_engfas* & \cmark &     Orig &       Orig &    0.578 & 0.444 & 0.493 &  0.758 & 0.915 \\
      hltcoe &                 plaid\_distill\_mono\_fasfas* & \xmark &     Orig &         GT &    0.574 & 0.453 & 0.513 &  0.747 & 0.912 \\
coordinators &                \ml{plaid\_distill\_mono\_ht*} & \cmark &     Orig &         HT &    0.567 & 0.440 & 0.510 &  0.714 & 0.887 \\
      hltcoe &            plaid\_distill\_engeng\_zs2engfas* & \cmark &     Orig &       Orig &    0.541 & 0.410 & 0.473 &  0.741 & 0.909 \\
      h2oloo &      mlir-ATNESLH-h2oloo-spladedt+rocchio-fas & \xmark &  Orig+DT &    Orig+GT &    0.515 & 0.407 & 0.428 &  0.705 & 0.907 \\
      h2oloo &                  mlir-ATNESLH-h2oloo-lit5-fas & \cmark &  Orig+DT &    Orig+GT &    0.499 & 0.410 & 0.438 &  0.763 & 0.936 \\
coordinators &                                  fast\_psqtd* & \cmark &     Orig &       Orig &    0.487 & 0.376 & 0.419 &  0.718 & 0.870 \\
      h2oloo &          mlir-ATESH-h2oloo-bm25dt+rocchio-fas & \xmark &  Orig+DT &    Orig+GT &    0.465 & 0.385 & 0.392 &  0.663 & 0.870 \\
coordinators &                        patapscoBM25dtnoRM3td* & \cmark &       DT &       Orig &    0.459 & 0.362 & 0.365 &  0.645 & 0.835 \\
coordinators &                          patapscoBM25qtRM3td* & \cmark &     Orig &         GT &    0.456 & 0.385 & 0.369 &  0.659 & 0.859 \\
coordinators &                          patapscoBM25dtRM3td* & \cmark &       DT &       Orig &    0.455 & 0.384 & 0.384 &  0.663 & 0.867 \\
coordinators &                        patapscoBM25qtnoRM3td* & \cmark &     Orig &         GT &    0.452 & 0.380 & 0.355 &  0.638 & 0.801 \\
coordinators &                   \ml{patapscoBM25htnoRM3td*} & \cmark &     Orig &         HT &    0.439 & 0.358 & 0.337 &  0.587 & 0.784 \\
coordinators &                               fast\_psqtitle* & \cmark &     Orig &       Orig &    0.438 & 0.337 & 0.371 &  0.645 & 0.854 \\
coordinators &                        patapscoBM25dtRM3desc* & \xmark &       DT &       Orig &    0.428 & 0.349 & 0.350 &  0.663 & 0.856 \\
coordinators &                       patapscoBM25dtRM3title* & \xmark &       DT &       Orig &    0.427 & 0.337 & 0.347 &  0.624 & 0.823 \\
coordinators &                     \ml{patapscoBM25htRM3td*} & \cmark &     Orig &         HT &    0.425 & 0.358 & 0.328 &  0.616 & 0.832 \\
coordinators &                     patapscoBM25dtnoRM3title* & \xmark &       DT &       Orig &    0.421 & 0.318 & 0.330 &  0.596 & 0.767 \\
coordinators &                       patapscoBM25qtRM3title* & \xmark &     Orig &         GT &    0.417 & 0.349 & 0.337 &  0.601 & 0.801 \\
coordinators &                      patapscoBM25dtnoRM3desc* & \xmark &       DT &       Orig &    0.411 & 0.325 & 0.329 &  0.624 & 0.796 \\
coordinators &                  \ml{patapscoBM25htRM3title*} & \cmark &     Orig &         HT &    0.396 & 0.326 & 0.304 &  0.571 & 0.787 \\
coordinators &                        patapscoBM25qtRM3desc* & \xmark &     Orig &         GT &    0.392 & 0.344 & 0.313 &  0.626 & 0.825 \\
coordinators &                     patapscoBM25qtnoRM3title* & \xmark &     Orig &         GT &    0.391 & 0.327 & 0.303 &  0.574 & 0.718 \\
coordinators &                      patapscoBM25qtnoRM3desc* & \xmark &     Orig &         GT &    0.387 & 0.328 & 0.300 &  0.576 & 0.761 \\
coordinators &                 \ml{patapscoBM25htnoRM3desc*} & \xmark &     Orig &         HT &    0.375 & 0.292 & 0.274 &  0.528 & 0.732 \\
coordinators &                \ml{patapscoBM25htnoRM3title*} & \xmark &     Orig &         HT &    0.373 & 0.312 & 0.287 &  0.534 & 0.706 \\
coordinators &                   \ml{patapscoBM25htRM3desc*} & \xmark &     Orig &         HT &    0.372 & 0.309 & 0.285 &  0.573 & 0.806 \\
\bottomrule
\end{tabular}

\end{table*}

\begin{table*}
\setlength\tabcolsep{0.5em}

\caption{Russian Results.
Monolingual runs, which use human translations of the queries, are marked as green.
* indicates manual runs.
Column ``JFD'' indicates whether the run is \underline{j}udged at \underline{f}ull \underline{d}epth, which is 100.
Other runs were judged to depth 50. 
}\label{tab:rus-full-results}
    \centering
\centering
\begin{tabular}{ll|ccc|ccccc}
\toprule
        Team &                                      Run Name &    JFD &     DS &         QS &  nDCG@20 &   RBP &    AP &  R@100 &  R@1k \\
\midrule
      hltcoe &  plaid\_distill\_engrus.mt5rerank.gpt4rerank* & \cmark &     Orig &       Orig &    0.593 & 0.553 & 0.532 &  0.784 & 0.968 \\
      hltcoe &       kitchen\_rankfuse.mt5rerank.gpt4rerank* & \cmark &  Orig+DT &    Orig+GT &    0.585 & 0.544 & 0.519 &  0.779 & 0.985 \\
      h2oloo &                  mlir-ATNESLH-h2oloo-rg4o-rus & \cmark &  Orig+DT &    Orig+GT &    0.584 & 0.521 & 0.516 &  0.820 & 0.960 \\
      h2oloo &               mlir-ATNESLH-h2oloo-rfusedb-rus & \cmark &  Orig+DT &    Orig+GT &    0.578 & 0.518 & 0.511 &  0.820 & 0.960 \\
      h2oloo &            mlir-ATNESLH-h2oloo-rl70b-rus.trec & \cmark &  Orig+DT &    Orig+GT &    0.564 & 0.512 & 0.494 &  0.820 & 0.960 \\
      h2oloo &                mlir-ATNESLH-h2oloo-rfused-rus & \cmark &  Orig+DT &    Orig+GT &    0.558 & 0.512 & 0.491 &  0.820 & 0.960 \\
      hltcoe &                 plaid\_distill\_engrus\_450p* & \xmark &     Orig &       Orig &    0.537 & 0.443 & 0.458 &  0.796 & 0.967 \\
         ISI &                   rus-ISI\_SEARCHER-ANE\_run1 & \cmark &     Orig &       Orig &    0.527 & 0.450 & 0.464 &  0.817 & 0.911 \\
      hltcoe &                            kitchen\_rankfuse* & \cmark &  Orig+DT &    Orig+GT &    0.526 & 0.434 & 0.447 &  0.803 & 0.985 \\
      h2oloo &               mlir-ATNESLH-h2oloo-rzephyr-rus & \cmark &  Orig+DT &    Orig+GT &    0.516 & 0.489 & 0.461 &  0.820 & 0.960 \\
      hltcoe &            plaid\_distill\_engeng\_zs2engrus* & \cmark &     Orig &       Orig &    0.514 & 0.427 & 0.413 &  0.747 & 0.949 \\
      h2oloo &           mlir-ATNESLH-h2oloo-monot5+lit5-rus & \cmark &  Orig+DT &    Orig+GT &    0.508 & 0.475 & 0.450 &  0.820 & 0.960 \\
      hltcoe &                  kitchen\_rankfuse,mt5rerank* & \cmark &  Orig+DT &    Orig+GT &    0.508 & 0.458 & 0.457 &  0.779 & 0.985 \\
      hltcoe &              plaid\_eqsynms\_distill\_engrus* & \cmark &     Orig &       Orig &    0.507 & 0.412 & 0.425 &  0.767 & 0.960 \\
      hltcoe &             plaid\_distill\_engrus.mt5rerank* & \cmark &     Orig &       Orig &    0.503 & 0.423 & 0.428 &  0.779 & 0.968 \\
      hltcoe &                       plaid\_distill\_engrus* & \cmark &     Orig &       Orig &    0.503 & 0.423 & 0.428 &  0.779 & 0.968 \\
      h2oloo &                mlir-ATNESLH-h2oloo-monot5-rus & \cmark &  Orig+DT &    Orig+GT &    0.502 & 0.448 & 0.445 &  0.782 & 0.960 \\
      h2oloo & mlir-ATNESLH-h2oloo-bm25dt+spladedt+plaid-rus & \cmark &  Orig+DT &    Orig+GT &    0.501 & 0.430 & 0.423 &  0.794 & 0.960 \\
      hltcoe &            plaid\_distill\_engrus\_termpool2* & \xmark &     Orig &       Orig &    0.501 & 0.416 & 0.420 &  0.783 & 0.963 \\
      hltcoe &                       plaid\_distill\_engeng* & \xmark &       DT &       Orig &    0.494 & 0.421 & 0.430 &  0.757 & 0.962 \\
      hltcoe &                      plaid\_distill\_engmlir* & \xmark &     Orig &       Orig &    0.493 & 0.415 & 0.422 &  0.762 & 0.961 \\
      hltcoe &                 plaid\_distill\_mono\_rusrus* & \xmark &     Orig &         GT &    0.490 & 0.414 & 0.416 &  0.787 & 0.952 \\
      hltcoe &                  plaid\_syn\_distill\_engrus* & \cmark &     Orig &       Orig &    0.489 & 0.390 & 0.398 &  0.757 & 0.954 \\
      hltcoe &            plaid\_distill\_engeng\_zs2rusrus* & \xmark &     Orig &         GT &    0.475 & 0.405 & 0.402 &  0.775 & 0.946 \\
coordinators &                \ml{plaid\_distill\_mono\_ht*} & \cmark &     Orig &         HT &    0.475 & 0.403 & 0.396 &  0.762 & 0.947 \\
      h2oloo &      mlir-ATNESLH-h2oloo-spladedt+rocchio-rus & \xmark &  Orig+DT &    Orig+GT &    0.446 & 0.398 & 0.387 &  0.677 & 0.894 \\
      h2oloo &                  mlir-ATNESLH-h2oloo-lit5-rus & \cmark &  Orig+DT &    Orig+GT &    0.423 & 0.420 & 0.375 &  0.763 & 0.960 \\
coordinators &                          patapscoBM25qtRM3td* & \cmark &     Orig &         GT &    0.419 & 0.377 & 0.338 &  0.670 & 0.895 \\
coordinators &                          patapscoBM25dtRM3td* & \cmark &       DT &       Orig &    0.399 & 0.353 & 0.322 &  0.634 & 0.876 \\
coordinators &                     \ml{patapscoBM25htRM3td*} & \cmark &     Orig &         HT &    0.395 & 0.349 & 0.324 &  0.630 & 0.863 \\
coordinators &                       patapscoBM25qtRM3title* & \xmark &     Orig &         GT &    0.395 & 0.345 & 0.309 &  0.651 & 0.893 \\
coordinators &                        patapscoBM25qtnoRM3td* & \cmark &     Orig &         GT &    0.378 & 0.340 & 0.305 &  0.635 & 0.859 \\
coordinators &                        patapscoBM25dtnoRM3td* & \cmark &       DT &       Orig &    0.367 & 0.311 & 0.283 &  0.642 & 0.860 \\
coordinators &                        patapscoBM25dtRM3desc* & \xmark &       DT &       Orig &    0.363 & 0.319 & 0.291 &  0.578 & 0.830 \\
coordinators &                                  fast\_psqtd* & \cmark &     Orig &       Orig &    0.359 & 0.306 & 0.289 &  0.626 & 0.842 \\
coordinators &                   \ml{patapscoBM25htnoRM3td*} & \cmark &     Orig &         HT &    0.358 & 0.315 & 0.283 &  0.601 & 0.835 \\
coordinators &                     patapscoBM25qtnoRM3title* & \xmark &     Orig &         GT &    0.352 & 0.317 & 0.277 &  0.588 & 0.824 \\
coordinators &                        patapscoBM25qtRM3desc* & \xmark &     Orig &         GT &    0.350 & 0.294 & 0.284 &  0.581 & 0.837 \\
coordinators &                  \ml{patapscoBM25htRM3title*} & \cmark &     Orig &         HT &    0.346 & 0.299 & 0.280 &  0.588 & 0.836 \\
coordinators &                       patapscoBM25dtRM3title* & \xmark &       DT &       Orig &    0.339 & 0.297 & 0.274 &  0.620 & 0.877 \\
coordinators &                   \ml{patapscoBM25htRM3desc*} & \xmark &     Orig &         HT &    0.337 & 0.292 & 0.280 &  0.554 & 0.807 \\
coordinators &                      patapscoBM25dtnoRM3desc* & \xmark &       DT &       Orig &    0.326 & 0.273 & 0.232 &  0.566 & 0.801 \\
coordinators &                \ml{patapscoBM25htnoRM3title*} & \xmark &     Orig &         HT &    0.310 & 0.284 & 0.253 &  0.535 & 0.784 \\
coordinators &                      patapscoBM25qtnoRM3desc* & \xmark &     Orig &         GT &    0.306 & 0.279 & 0.244 &  0.542 & 0.779 \\
coordinators &                 \ml{patapscoBM25htnoRM3desc*} & \xmark &     Orig &         HT &    0.304 & 0.270 & 0.244 &  0.543 & 0.768 \\
coordinators &                     patapscoBM25dtnoRM3title* & \xmark &       DT &       Orig &    0.299 & 0.272 & 0.235 &  0.579 & 0.820 \\
coordinators &                               fast\_psqtitle* & \xmark &     Orig &       Orig &    0.296 & 0.247 & 0.235 &  0.539 & 0.797 \\
      h2oloo &          mlir-ATESH-h2oloo-bm25dt+rocchio-rus & \xmark &  Orig+DT &    Orig+GT &    0.000 & 0.000 & 0.000 &  0.000 & 0.000 \\
\bottomrule
\end{tabular}

\end{table*}

\begin{table*}
\setlength\tabcolsep{0.5em}

\caption{MLIR Results.
The run used as the first stage retrieval for the reranking task is marked in bold. 
* indicates manual runs.
\textbf{Results may change after TREC conference.}
Column ``JFD'' indicates whether the run is \underline{j}udged at \underline{f}ull \underline{d}epth, which is 100, otherwise 50. 
}\label{tab:mlir-full-results}
    \centering

\resizebox{\linewidth}{!}{
\begin{tabular}{ll|ccc|ccccc}
\toprule
           Team &                                                Run Name &    JFD &    DS &         QS &  nDCG@20 &   RBP &    AP &  R@100 &  R@1k \\
\midrule
         h2oloo &        mlir-ATNESLH-h2oloo-rfusedbinterweave-rlp2\_clir & \cmark &  Orig+DT &    Orig+GT &    0.545 & 0.590 & 0.460 &  0.702 & 0.917 \\
         h2oloo &        mlir-ATNESLH-h2oloo-rfusedbinterweave-rlp1\_clir & \cmark &  Orig+DT &    Orig+GT &    0.544 & 0.595 & 0.450 &  0.694 & 0.917 \\
         h2oloo & mlir-ATNESLH-h2oloo-rfusedbinterweave-rlp2\_rg4of\_clir & \cmark &  Orig+DT &    Orig+GT &    0.544 & 0.603 & 0.467 &  0.702 & 0.917 \\
         h2oloo &  mlir-ATNESLH-h2oloo-rfusedbinterweave-rlp2\_rg4o\_clir & \cmark &  Orig+DT &    Orig+GT &    0.540 & 0.604 & 0.472 &  0.702 & 0.917 \\
         hltcoe &       kitchen\_rankfuse.mt5rerank.scorefuse.gpt4rerank* & \cmark &  Orig+DT &    Orig+GT &    0.521 & 0.630 & 0.457 &  0.663 & 0.903 \\
         hltcoe &    plaid\_distill\_clir.mt5rerank.scorefuse.gpt4rerank* & \cmark &     Orig &       Orig &    0.520 & 0.625 & 0.442 &  0.656 & 0.878 \\
         h2oloo &              mlir-ATNESLH-h2oloo-rfusedbinterweave-clir & \cmark &  Orig+DT &    Orig+GT &    0.511 & 0.564 & 0.414 &  0.678 & 0.917 \\
            ISI &                            mlir-ISI\_SEARCHER-ANE\_run1 & \cmark &     Orig &       Orig &    0.490 & 0.568 & 0.468 &  0.736 & 0.844 \\
         hltcoe &                  kitchen\_rankfuse.mt5rerank.scorefuse* & \xmark &  Orig+DT &    Orig+GT &    0.475 & 0.562 & 0.438 &  0.663 & 0.903 \\
         hltcoe &               plaid\_distill\_clir.mt5rerank.scorefuse* & \xmark &     Orig &       Orig &    0.474 & 0.560 & 0.431 &  0.656 & 0.878 \\
         hltcoe &                        plaid\_distill\_mlir\_mixedpass* & \cmark &     Orig &       Orig &    0.468 & 0.502 & 0.416 &  0.656 & 0.889 \\
         hltcoe &            plaid\_distill\_mlir\_mixedentry\_termpool2* & \cmark &     Orig &       Orig &    0.462 & 0.510 & 0.420 &  0.652 & 0.886 \\
         hltcoe &                                 plaid\_distill\_engeng* & \cmark &       DT &       Orig &    0.453 & 0.524 & 0.419 &  0.678 & 0.887 \\
         hltcoe &                       plaid\_distill\_mlir\_mixedentry* & \cmark &     Orig &       Orig &    0.444 & 0.494 & 0.402 &  0.654 & 0.890 \\
         hltcoe &                plaid\_distill\_mlir\_bycoll\_scorefuse* & \cmark &     Orig &       Orig &    0.438 & 0.493 & 0.389 &  0.639 & 0.894 \\
         hltcoe &                               plaid\_distill\_mlir\_rr* & \cmark &     Orig &       Orig &    0.428 & 0.479 & 0.383 &  0.642 & 0.886 \\
IRLab-Amsterdam &                      mlir-IRLabAmsterdam-ANEL-titledesc & \cmark &     Orig &       Orig &    0.354 & 0.415 & 0.303 &  0.554 & 0.818 \\
         hltcoe &                        plaid\_distill\_clir\_scorefuse* & \xmark &     Orig &       Orig &    0.353 & 0.431 & 0.315 &  0.577 & 0.827 \\
IRLab-Amsterdam &                          mlir-IRLabAmsterdam-ANEL-title & \cmark &     Orig &       Orig &    0.352 & 0.414 & 0.285 &  0.485 & 0.754 \\
   coordinators &                                    patapscoBM25dtRM3td* & \cmark &       DT &       Orig &    0.350 & 0.431 & 0.283 &  0.507 & 0.827 \\
         hltcoe &                             plaid\_distill\_engeng\_zs* & \cmark &     Orig &       Orig &    0.343 & 0.392 & 0.304 &  0.568 & 0.838 \\
   coordinators &                                  patapscoBM25dtnoRM3td* & \cmark &       DT &       Orig &    0.337 & 0.402 & 0.239 &  0.500 & 0.753 \\
IRLab-Amsterdam &                           mlir-IRLabAmsterdam-ANEL-desc & \cmark &     Orig &       Orig &    0.335 & 0.388 & 0.276 &  0.516 & 0.800 \\
   coordinators &                                            fast\_psqtd* & \cmark &     Orig &       Orig &    0.322 & 0.401 & 0.273 &  0.516 & 0.765 \\
   coordinators &                                 patapscoBM25dtRM3title* & \cmark &       DT &       Orig &    0.315 & 0.389 & 0.253 &  0.466 & 0.772 \\
   coordinators &                                  patapscoBM25dtRM3desc* & \cmark &       DT &       Orig &    0.306 & 0.374 & 0.243 &  0.458 & 0.790 \\
   coordinators &                               patapscoBM25dtnoRM3title* & \cmark &       DT &       Orig &    0.298 & 0.354 & 0.213 &  0.439 & 0.710 \\
   coordinators &                                patapscoBM25dtnoRM3desc* & \cmark &       DT &       Orig &    0.279 & 0.339 & 0.195 &  0.458 & 0.705 \\
   coordinators &                                         fast\_psqtitle* & \cmark &     Orig &       Orig &    0.263 & 0.307 & 0.205 &  0.455 & 0.716 \\

\bottomrule
\end{tabular}
}

\end{table*}

\begin{table*}
\caption{Technical Document Task Results.
Monolingual runs, which use human translations of the queries, are shown in green.
* indicates manual runs. All runs are judged to depth 40. 
}\label{tab:tech-full-results}
    \centering
\begin{tabular}{ll|cc|ccccc}
\toprule
        Team &                                     Run Name &       DS &         QS &  nDCG@20 &   RBP &    AP &  R@100 &  R@1k \\
\midrule
         ISI &                 tech-ISI\_SEARCHER-ANE\_run1 &     Orig &       Orig &    0.496 & 0.468 & 0.350 &  0.638 & 0.807 \\
      h2oloo &                      fs1\_monot5\_listgalore &  Orig+DT &    Orig+GT &    0.491 & 0.481 & 0.355 &  0.573 & 0.837 \\
      h2oloo &                fs1\_xlmr\_monot5\_listgalore &  Orig+DT &    Orig+GT &    0.489 & 0.485 & 0.370 &  0.592 & 0.937 \\
      h2oloo &                   fs1\_xlmr\_monot5\_rgpt-4o &  Orig+DT &    Orig+GT &    0.488 & 0.485 & 0.372 &  0.592 & 0.937 \\
      h2oloo &                         fs1\_monot5\_rgpt-4o &  Orig+DT &    Orig+GT &    0.482 & 0.473 & 0.353 &  0.573 & 0.837 \\
      h2oloo &                      fs1\_monot5\_rl3.1\_70b &  Orig+DT &    Orig+GT &    0.477 & 0.473 & 0.345 &  0.573 & 0.837 \\
      h2oloo &                fs1\_xlmr\_monot5\_rl3.1\_70b &  Orig+DT &    Orig+GT &    0.474 & 0.467 & 0.353 &  0.592 & 0.937 \\
      hltcoe & plaid\_distill\_engzho.mt5rerank.gpt4rerank* &     Orig &       Orig &    0.460 & 0.469 & 0.340 &  0.614 & 0.842 \\
      hltcoe &      kitchen\_rankfuse.mt5rerank.gpt4rerank* &  Orig+DT &    Orig+GT &    0.459 & 0.464 & 0.340 &  0.644 & 0.922 \\
      h2oloo &                              fs1\_monot5\_rz &  Orig+DT &    Orig+GT &    0.454 & 0.440 & 0.318 &  0.573 & 0.837 \\
      h2oloo &                        fs1\_xlmr\_monot5\_rz &  Orig+DT &    Orig+GT &    0.448 & 0.449 & 0.328 &  0.592 & 0.937 \\
      hltcoe &                 kitchen\_rankfuse.mt5rerank* &  Orig+DT &    Orig+GT &    0.440 & 0.431 & 0.327 &  0.644 & 0.922 \\
      hltcoe &            plaid\_distill\_engzho.mt5rerank* &     Orig &       Orig &    0.434 & 0.430 & 0.320 &  0.614 & 0.842 \\
coordinators &                                  topic\_dev* &  Orig+DT &      Other &    0.407 & 0.462 & 0.227 &  0.267 & 0.267 \\
coordinators &               \ml{plaid\_distill\_mono\_ht*} &     Orig &         HT &    0.406 & 0.403 & 0.284 &  0.593 & 0.848 \\
      h2oloo &                                  fs1\_monot5 &  Orig+DT &    Orig+GT &    0.397 & 0.379 & 0.276 &  0.573 & 0.837 \\
      h2oloo &                            fs1\_xlmr\_monot5 &  Orig+DT &    Orig+GT &    0.393 & 0.383 & 0.287 &  0.592 & 0.937 \\
      hltcoe &             plaid\_eqsynms\_distill\_engzho* &     Orig &       Orig &    0.389 & 0.389 & 0.291 &  0.578 & 0.872 \\
      h2oloo &                                    fs1\_xlmr &  Orig+DT &    Orig+GT &    0.388 & 0.391 & 0.285 &  0.628 & 0.937 \\
      hltcoe &           plaid\_distill\_engeng\_zs2zhozho* &     Orig &         GT &    0.384 & 0.369 & 0.267 &  0.568 & 0.857 \\
      hltcoe &                           kitchen\_rankfuse* &  Orig+DT &    Orig+GT &    0.383 & 0.365 & 0.269 &  0.580 & 0.922 \\
      hltcoe &                      plaid\_distill\_zhozho* &     Orig &         GT &    0.377 & 0.373 & 0.265 &  0.558 & 0.851 \\
      hltcoe &                      plaid\_distill\_engeng* &       DT &       Orig &    0.373 & 0.376 & 0.264 &  0.534 & 0.838 \\
      hltcoe &                      plaid\_distill\_engzho* &     Orig &       Orig &    0.372 & 0.367 & 0.268 &  0.563 & 0.842 \\
      hltcoe &                 plaid\_syn\_distill\_engzho* &     Orig &       Orig &    0.361 & 0.361 & 0.264 &  0.541 & 0.845 \\
      h2oloo &                                          fs1 &  Orig+DT &    Orig+GT &    0.309 & 0.305 & 0.216 &  0.531 & 0.837 \\
coordinators &                  \ml{patapscoBM25htnoRM3td*} &     Orig &         HT &    0.309 & 0.307 & 0.187 &  0.418 & 0.651 \\
coordinators &                 \ml{patapscoBM25htRM3title*} &     Orig &         HT &    0.293 & 0.291 & 0.197 &  0.418 & 0.660 \\
coordinators &                                fast\_psq\_t* &     Orig &       Orig &    0.280 & 0.274 & 0.164 &  0.365 & 0.656 \\
coordinators &               \ml{patapscoBM25htnoRM3title*} &     Orig &         HT &    0.277 & 0.283 & 0.175 &  0.388 & 0.629 \\
coordinators &                    \ml{patapscoBM25htRM3td*} &     Orig &         HT &    0.276 & 0.280 & 0.178 &  0.397 & 0.691 \\
coordinators &                    patapscoBM25dtnoRM3title* &       DT &       Orig &    0.275 & 0.285 & 0.191 &  0.426 & 0.668 \\
coordinators &                      patapscoBM25dtRM3title* &       DT &       Orig &    0.275 & 0.285 & 0.191 &  0.426 & 0.668 \\
coordinators &                               fast\_psq\_td* &     Orig &       Orig &    0.272 & 0.267 & 0.163 &  0.387 & 0.661 \\
      hltcoe &           plaid\_distill\_engeng\_zs2engzho* &     Orig &       Orig &    0.268 & 0.272 & 0.191 &  0.455 & 0.740 \\
coordinators &                       patapscoBM25dtnoRM3td* &       DT &       Orig &    0.266 & 0.268 & 0.166 &  0.397 & 0.653 \\
coordinators &                       patapscoBM25qtnoRM3td* &     Orig &         GT &    0.257 & 0.253 & 0.151 &  0.380 & 0.629 \\
coordinators &                  \ml{patapscoBM25htRM3desc*} &     Orig &         HT &    0.255 & 0.267 & 0.164 &  0.372 & 0.635 \\
      h2oloo &                   bm25-rocchio-qt-desc+title &     Orig &         GT &    0.253 & 0.248 & 0.175 &  0.398 & 0.715 \\
coordinators &                         patapscoBM25qtRM3td* &     Orig &         GT &    0.253 & 0.259 & 0.160 &  0.355 & 0.652 \\
coordinators &                \ml{patapscoBM25htnoRM3desc*} &     Orig &         HT &    0.252 & 0.260 & 0.150 &  0.366 & 0.586 \\
coordinators &                         patapscoBM25dtRM3td* &       DT &       Orig &    0.252 & 0.245 & 0.165 &  0.402 & 0.665 \\
coordinators &                      patapscoBM25qtRM3title* &     Orig &         GT &    0.239 & 0.250 & 0.155 &  0.354 & 0.615 \\
coordinators &                    patapscoBM25qtnoRM3title* &     Orig &         GT &    0.230 & 0.240 & 0.142 &  0.355 & 0.580 \\
      h2oloo &                   bm25-rocchio-dt-desc+title &       DT &       Orig &    0.222 & 0.222 & 0.146 &  0.378 & 0.656 \\
coordinators &                     patapscoBM25dtnoRM3desc* &       DT &       Orig &    0.217 & 0.223 & 0.131 &  0.340 & 0.590 \\
coordinators &                       patapscoBM25dtRM3desc* &       DT &       Orig &    0.212 & 0.215 & 0.148 &  0.365 & 0.591 \\
coordinators &                       patapscoBM25qtRM3desc* &     Orig &         GT &    0.211 & 0.214 & 0.128 &  0.315 & 0.583 \\
coordinators &                     patapscoBM25qtnoRM3desc* &     Orig &         GT &    0.211 & 0.212 & 0.116 &  0.311 & 0.550 \\
      h2oloo &                          gte-qwen-desc+title &     Orig &         GT &    0.195 & 0.193 & 0.122 &  0.338 & 0.629 \\
\bottomrule
\end{tabular}
\begin{flushleft}
\end{flushleft}

\end{table*}

\begin{table*}[]
\caption{Chinese Report Generation Task Scores. Values in parentheses are standard deviations across topics. }\label{tab:zho-repgen-results}
\centering

\begin{tabular}{ll|c|cccc}
\toprule
{}        & {}                                               & ARGUE         & Citation      & Nugget        & Nugget        & Sentence      \\
Team      & Run ID                                           & Score         & Precision     & Recall        & Support       & Support       \\
\midrule

hltcoe    & zho-hltcoe-eugene-gpt4o-fixed                    & 0.726 (0.263) & 0.859 (0.266) & 0.327 (0.219) & 0.298 (0.169) & 0.840 (0.136) \\
IDA       & IDA\_CCS\_hybrid\_zho                            & 0.637 (0.310) & 0.795 (0.261) & 0.236 (0.164) & 0.166 (0.120) & 0.803 (0.234) \\
hltcoe    & zho-hltcoe-eugene-gpt35turbo                     & 0.587 (0.284) & 0.813 (0.291) & 0.250 (0.160) & 0.248 (0.171) & 0.720 (0.190) \\
hltcoe    & zho-jhu-orion-aggregated-w-claude                & 0.528 (0.251) & 0.900 (0.249) & 0.177 (0.123) & 0.238 (0.201) & 0.662 (0.200) \\
hltcoe    & zho-jhu-orion-aggregated-w-gpt4o                 & 0.496 (0.277) & 0.899 (0.234) & 0.182 (0.145) & 0.237 (0.184) & 0.636 (0.226) \\
irlab-ams & zho\_irlab-ams-std-translate-llama-70B-api       & 0.464 (0.295) & 0.927 (0.174) & 0.181 (0.123) & 0.243 (0.150) & 0.574 (0.227) \\
IDA       & IDA\_CCS\_abstractive\_zho                       & 0.456 (0.226) & 0.873 (0.187) & 0.230 (0.144) & 0.184 (0.132) & 0.601 (0.206) \\
h2oloo    & rfused\_rgn\_crp\_zho                            & 0.432 (0.206) & 0.895 (0.190) & 0.261 (0.138) & 0.292 (0.182) & 0.555 (0.152) \\
h2oloo    & rfused\_rgn\_l70b\_zho                           & 0.376 (0.246) & 0.850 (0.246) & 0.236 (0.187) & 0.249 (0.172) & 0.433 (0.223) \\
irlab-ams & zho\_irlab-ams-std-translate-llama-8B            & 0.360 (0.171) & 0.861 (0.196) & 0.209 (0.157) & 0.164 (0.112) & 0.563 (0.193) \\
h2oloo    & rfused\_rgn\_gpt4o\_zho                          & 0.348 (0.181) & 0.894 (0.216) & 0.246 (0.150) & 0.171 (0.132) & 0.594 (0.180) \\
h2oloo    & rfused\_rgn\_l70bph\_zho                         & 0.346 (0.212) & 0.894 (0.233) & 0.260 (0.185) & 0.286 (0.181) & 0.399 (0.225) \\
irlab-ams & zho\_irlab-ams-std-recomp-llama-8B               & 0.277 (0.177) & 0.777 (0.227) & 0.158 (0.119) & 0.164 (0.174) & 0.384 (0.181) \\
irlab-ams & zho\_irlab-ams-postcite                          & 0.181 (0.241) & 0.546 (0.297) & 0.102 (0.137) & 0.135 (0.187) & 0.229 (0.235) \\
irlab-ams & zho\_irlab-ams-std-mdcomp-330-translate-llama-8B & 0.166 (0.138) & 0.759 (0.271) & 0.135 (0.113) & 0.104 (0.086) & 0.269 (0.174) \\
irlab-ams & zho\_irlab-ams-std-mdcomp-331-translate-llama-8B & 0.144 (0.108) & 0.744 (0.270) & 0.151 (0.103) & 0.089 (0.061) & 0.236 (0.145) \\
irlab-ams & zho\_irlab-ams-postcite-v                        & 0.121 (0.140) & 0.423 (0.293) & 0.105 (0.131) & 0.115 (0.143) & 0.156 (0.159) \\
\bottomrule
\end{tabular}

\end{table*}

\begin{table*}[]
\caption{Persian Report Generation Task Scores. Values in parentheses are standard deviations across topics. }\label{tab:fas-repgen-results}
\centering

\begin{tabular}{ll|c|cccc}
\toprule
{}        & {}                                               & ARGUE         & Citation      & Nugget        & Nugget        & Sentence      \\
Team      & Run ID                                           & Score         & Precision     & Recall        & Support       & Support       \\
\midrule
hltcoe    & fas-hltcoe-eugene-gpt4o                          & 0.872 (0.078) & 0.918 (0.115) & 0.303 (0.167) & 0.311 (0.134) & 0.919 (0.061) \\
IDA       & IDA\_CCS\_hybrid\_fas                            & 0.681 (0.320) & 0.853 (0.149) & 0.271 (0.197) & 0.183 (0.139) & 0.845 (0.193) \\
hltcoe    & fas-hltcoe-eugene-gpt35turbo                     & 0.646 (0.290) & 0.846 (0.215) & 0.215 (0.162) & 0.201 (0.136) & 0.792 (0.202) \\
irlab-ams & fas\_irlab-ams-std-translate-llama-70B-api       & 0.566 (0.261) & 0.852 (0.211) & 0.197 (0.127) & 0.272 (0.122) & 0.710 (0.270) \\
IDA       & IDA\_CCS\_abstractive\_fas                       & 0.525 (0.271) & 0.940 (0.106) & 0.282 (0.182) & 0.215 (0.140) & 0.621 (0.252) \\
hltcoe    & fas-jhu-orion-aggregated-w-claude                & 0.519 (0.173) & 0.945 (0.121) & 0.213 (0.114) & 0.268 (0.156) & 0.650 (0.174) \\
h2oloo    & rfused\_rgn\_l70b\_fas                           & 0.463 (0.234) & 0.931 (0.097) & 0.231 (0.163) & 0.270 (0.162) & 0.551 (0.222) \\
hltcoe    & fas-jhu-orion-aggregated-w-gpt4o                 & 0.449 (0.265) & 0.941 (0.089) & 0.220 (0.165) & 0.272 (0.182) & 0.565 (0.243) \\
h2oloo    & rfused\_rgn\_crp\_fas                            & 0.431 (0.224) & 0.889 (0.169) & 0.295 (0.197) & 0.253 (0.156) & 0.586 (0.169) \\
h2oloo    & rfused\_rgn\_l70bph\_fas                         & 0.420 (0.224) & 0.940 (0.129) & 0.240 (0.147) & 0.328 (0.208) & 0.504 (0.185) \\
irlab-ams & fas\_irlab-ams-std-translate-llama-8B            & 0.337 (0.219) & 0.795 (0.181) & 0.182 (0.130) & 0.183 (0.135) & 0.553 (0.246) \\
h2oloo    & rfused\_rgn\_gpt4o\_fas                          & 0.304 (0.162) & 0.934 (0.153) & 0.230 (0.149) & 0.189 (0.119) & 0.572 (0.207) \\
irlab-ams & fas\_irlab-ams-std-recomp-llama-8B               & 0.292 (0.220) & 0.781 (0.216) & 0.133 (0.137) & 0.155 (0.165) & 0.412 (0.225) \\
irlab-ams & fas\_irlab-ams-postcite                          & 0.188 (0.189) & 0.452 (0.253) & 0.108 (0.167) & 0.105 (0.129) & 0.266 (0.238) \\
irlab-ams & fas\_irlab-ams-std-mdcomp-330-translate-llama-8B & 0.179 (0.105) & 0.703 (0.244) & 0.171 (0.173) & 0.118 (0.083) & 0.290 (0.101) \\
irlab-ams & fas\_irlab-ams-std-mdcomp-331-translate-llama-8B & 0.159 (0.159) & 0.665 (0.282) & 0.150 (0.138) & 0.107 (0.116) & 0.257 (0.181) \\
irlab-ams & fas\_irlab-ams-postcite-v                        & 0.087 (0.114) & 0.389 (0.262) & 0.058 (0.083) & 0.058 (0.086) & 0.147 (0.147) \\
\bottomrule
\end{tabular}

\end{table*}

\begin{table*}[]
\caption{Russian Report Generation Task Scores. Values in parentheses are standard deviations across topics. }\label{tab:rus-repgen-results}
\centering

\begin{tabular}{ll|c|cccc}
\toprule
{}        & {}                                               & ARGUE         & Citation      & Nugget        & Nugget        & Sentence      \\
Team      & Run ID                                           & Score         & Precision     & Recall        & Support       & Support       \\
\midrule
hltcoe    & rus-hltcoe-eugene-gpt4o                          & 0.808 (0.208) & 0.904 (0.133) & 0.339 (0.167) & 0.420 (0.192) & 0.874 (0.139) \\
IDA       & IDA\_CCS\_hybrid\_rus                            & 0.615 (0.360) & 0.768 (0.288) & 0.296 (0.221) & 0.235 (0.150) & 0.799 (0.212) \\
hltcoe    & rus-hltcoe-eugene-gpt35turbo                     & 0.602 (0.305) & 0.799 (0.293) & 0.313 (0.207) & 0.309 (0.198) & 0.715 (0.242) \\
hltcoe    & rus-jhu-orion-aggregated-w-claude                & 0.519 (0.284) & 0.902 (0.208) & 0.255 (0.161) & 0.323 (0.189) & 0.673 (0.274) \\
irlab-ams & rus\_irlab-ams-std-translate-llama-70B-api       & 0.472 (0.229) & 0.871 (0.209) & 0.255 (0.169) & 0.393 (0.243) & 0.566 (0.231) \\
h2oloo    & rfused\_rgn\_l70b\_rus                           & 0.469 (0.246) & 0.903 (0.198) & 0.278 (0.203) & 0.319 (0.212) & 0.527 (0.270) \\
hltcoe    & rus-jhu-orion-aggregated-w-gpt4o                 & 0.415 (0.267) & 0.904 (0.198) & 0.247 (0.165) & 0.287 (0.187) & 0.495 (0.278) \\
h2oloo    & rfused\_rgn\_crp\_rus                            & 0.414 (0.223) & 0.914 (0.185) & 0.293 (0.209) & 0.304 (0.193) & 0.498 (0.220) \\
IDA       & IDA\_CCS\_abstractive\_rus                       & 0.403 (0.296) & 0.851 (0.241) & 0.298 (0.227) & 0.235 (0.197) & 0.486 (0.297) \\
h2oloo    & rfused\_rgn\_l70bph\_rus                         & 0.403 (0.202) & 0.894 (0.210) & 0.298 (0.193) & 0.363 (0.237) & 0.469 (0.220) \\
h2oloo    & rfused\_rgn\_gpt4o\_rus                          & 0.309 (0.174) & 0.927 (0.139) & 0.284 (0.170) & 0.203 (0.155) & 0.536 (0.238) \\
irlab-ams & rus\_irlab-ams-std-translate-llama-8B            & 0.265 (0.190) & 0.889 (0.178) & 0.216 (0.162) & 0.188 (0.122) & 0.436 (0.221) \\
irlab-ams & rus\_irlab-ams-std-recomp-llama-8B               & 0.227 (0.194) & 0.719 (0.261) & 0.152 (0.141) & 0.165 (0.137) & 0.380 (0.227) \\
irlab-ams & rus\_irlab-ams-std-mdcomp-330-translate-llama-8B & 0.185 (0.149) & 0.737 (0.264) & 0.157 (0.147) & 0.164 (0.136) & 0.264 (0.168) \\
irlab-ams & rus\_irlab-ams-std-mdcomp-331-translate-llama-8B & 0.185 (0.134) & 0.739 (0.269) & 0.168 (0.111) & 0.153 (0.115) & 0.292 (0.145) \\
irlab-ams & rus\_irlab-ams-postcite                          & 0.126 (0.175) & 0.480 (0.230) & 0.085 (0.136) & 0.084 (0.118) & 0.216 (0.200) \\
irlab-ams & rus\_irlab-ams-postcite-v                        & 0.074 (0.122) & 0.439 (0.245) & 0.050 (0.091) & 0.070 (0.095) & 0.091 (0.120) \\

\bottomrule
\end{tabular}

\end{table*}

\end{document}